\documentclass[
twocolumn,
frontmatterverbose,
nofootinbib,
noeprint,
longbibliography,
amsmath,
amssymb,
prd,
colorlinks,
hyperindex,
breaklinks,
linkcolor=cyan,
citecolor=cyan,
urlcolor=cyan,
]{revtex4-2}

\usepackage{graphicx}
\usepackage{dcolumn}
\usepackage{bm}
\usepackage{float}
\usepackage{xcolor}
\usepackage{multirow}
\usepackage{xspace}
\usepackage{orcidlink}
\usepackage{mathtools}
\usepackage[normalem]{ulem}

\def\belletwo {Belle~II}

\def\nub        {\ensuremath{\overline{\nu}}\xspace}

\def\nub        {\ensuremath{\overline{\nu}}\xspace}

\def\qqbar {\ensuremath{q\overline q}\xspace}

\def\ssbar {\ensuremath{s\overline s}\xspace}

\def\Kbar  {\kern 0.2em\overline{\kern -0.2em K}{}\xspace}

\def\Kz    {\ensuremath{K^0}\xspace}
\def\Kzb   {\ensuremath{\Kbar^0}\xspace}
\def\KzKzb {\ensuremath{\Kz \kern -0.16em \Kzb}\xspace}
\def\Kp    {\ensuremath{K^+}\xspace}
\def\Km    {\ensuremath{K^-}\xspace}

\def\KpKm  {\ensuremath{\Kp \kern -0.16em \Km}\xspace}
\def\KS    {\ensuremath{K^0_{\scriptscriptstyle S}}\xspace} 
\def\KL    {\ensuremath{K^0_{\scriptscriptstyle L}}\xspace}

\def\Dbar    {\kern 0.2em\overline{\kern -0.2em D}{}\xspace}

\def\Dz      {\ensuremath{D^0}\xspace}
\def\Dzb     {\ensuremath{\Dbar^0}\xspace}
\def\DzDzb   {\ensuremath{\Dz {\kern -0.16em \Dzb}}\xspace}
\def\Dp      {\ensuremath{D^+}\xspace}
\def\Dm      {\ensuremath{D^-}\xspace}

\def\DpDm    {\ensuremath{\Dp {\kern -0.16em \Dm}}\xspace}

\def\B       {\ensuremath{B}\xspace}
\def\Bbar    {\kern 0.18em\overline{\kern -0.18em B}{}\xspace}

\def\BB      {\ensuremath{B\Bbar}\xspace} 
\def\Bz      {\ensuremath{B^0}\xspace}
\def\Bzb     {\ensuremath{\Bbar^0}\xspace}
\def\BzBzb   {\ensuremath{\Bz {\kern -0.16em \Bzb}}\xspace}
\def\Bu      {\ensuremath{B^+}\xspace}
\def\Bub     {\ensuremath{B^-}\xspace}

\def\BpBm    {\ensuremath{\Bu {\kern -0.16em \Bub}}\xspace}

\def\jpsi     {\ensuremath{{J\mskip -3mu/\mskip -2mu\psi\mskip 2mu}}\xspace}

\mathchardef\Upsilon="7107
\def\Y#1S{\ensuremath{\Upsilon{(#1S)}}\xspace}

\mathchardef\Deltares="7101
\mathchardef\Xi="7104
\mathchardef\Lambda="7103
\mathchardef\Sigma="7106
\mathchardef\Omega="710A

\def\Deltabar{\kern 0.25em\overline{\kern -0.25em \Deltares}{}\xspace}
\def\Lbar{\kern 0.2em\overline{\kern -0.2em\Lambda\kern 0.05em}\kern-0.05em{}\xspace}
\def\Sigbar{\kern 0.2em\overline{\kern -0.2em \Sigma}{}\xspace}
\def\Xibar{\kern 0.2em\overline{\kern -0.2em \Xi}{}\xspace}
\def\Obar{\kern 0.2em\overline{\kern -0.2em \Omega}{}\xspace}
\def\Nbar{\kern 0.2em\overline{\kern -0.2em N}{}\xspace}
\def\Xb{\kern 0.2em\overline{\kern -0.2em X}{}\xspace}

\def\X {\ensuremath{X}\xspace}

\newcommand{\tev}{\ensuremath{\mathrm{\,Te\kern -0.1em V}}\xspace}
\newcommand{\gev}{\ensuremath{\mathrm{\,Ge\kern -0.1em V}}\xspace}
\newcommand{\mev}{\ensuremath{\mathrm{\,Me\kern -0.1em V}}\xspace}
\newcommand{\kev}{\ensuremath{\mathrm{\,ke\kern -0.1em V}}\xspace}
\newcommand{\ev}{\ensuremath{\mathrm{\,e\kern -0.1em V}}\xspace}
\newcommand{\gevc}{\ensuremath{{\mathrm{\,Ge\kern -0.1em V\!/}c}}\xspace}
\newcommand{\mevc}{\ensuremath{{\mathrm{\,Me\kern -0.1em V\!/}c}}\xspace}
\newcommand{\gevcc}{\ensuremath{{\mathrm{\,Ge\kern -0.1em V\!/}c^2}}\xspace}
\newcommand{\mevcc}{\ensuremath{{\mathrm{\,Me\kern -0.1em V\!/}c^2}}\xspace}

\newcommand{\gevvcc}{\ensuremath{{\mathrm{\,Ge\kern -0.1em V^2\!/}c^2}}\xspace}
\newcommand{\gevv}{\ensuremath{\mathrm{\,Ge\kern -0.1em V^2}}\xspace}
\newcommand{\gevvv}{\ensuremath{\mathrm{\,Ge\kern -0.1em V^3}}\xspace}
\newcommand{\gevvcccc}{\ensuremath{{\mathrm{\,Ge\kern -0.1em V^2\!/}c^4}}\xspace}

\def\cm   {\ensuremath{{\rm \,cm}}\xspace}

\def\invfb   {\ensuremath{\mbox{\,fb}^{-1}}\xspace}

\def\mus  {\ensuremath{\rm \,\mus}\xspace}

\def\ps   {\ensuremath{\rm \,ps}\xspace}

\def\mus        {\ensuremath{\,\mu{\rm s}}\xspace}    
\def\ps         {\ensuremath{{\rm \,ps}}\xspace}  

\def\to                 {\ensuremath{\rightarrow}\xspace}

\def\gsim{{~\raise.15em\hbox{$>$}\kern-.85em
          \lower.35em\hbox{$\sim$}~}\xspace}
\def\lsim{{~\raise.15em\hbox{$<$}\kern-.85em
          \lower.35em\hbox{$\sim$}~}\xspace}

\newcommand{\Vub}{\ensuremath{V_{ub}}\xspace}

\newcommand{\Vcb}{\ensuremath{V_{cb}}\xspace}

\def\BES {\text{BES{\ }\uppercase\expandafter{\romannumeral3}}}
\def\BTWO {\text{Belle{\ }\uppercase\expandafter{\romannumeral2}}}
\def\belle {\text{Belle{\ }}}

\newcommand{\BXulnu}{\ensuremath{\Bbar \to X_u\ell\nub}\xspace}

\newcommand{\BXclnu}{\ensuremath{\Bbar\to X_c\ell\nub}\xspace}

\newcommand{\lepp} {\ensuremath{E_\ell^{B_{\textrm{sig}}}}\xspace}

\newcommand{\BXuclnu}{\ensuremath{\Bbar\to X_{u/c}\ell\nub}\xspace}
\newcommand{\Xulnu}{\ensuremath{X_{u}\ell\nub}\xspace}
\newcommand{\Xclnu}{\ensuremath{X_{c}\ell\nub}\xspace}

\allowdisplaybreaks
\begin{document}

\title{Measurement of the Ratio of Partial Branching Fractions of Inclusive \texorpdfstring{$\Bbar \to X_{u} \ell \nub$ to $\Bbar \to X_{c} \ell \nub$ }{Semileptonic Charmless to Charmed B Decays} and the Ratio of their Spectra with Hadronic Tagging}

\noaffiliation
  \author{M.~Hohmann\,\orcidlink{0000-0001-5147-4781}} 
  \author{P.~Urquijo\,\orcidlink{0000-0002-0887-7953}} 
  \author{I.~Adachi\,\orcidlink{0000-0003-2287-0173}} 
  \author{H.~Aihara\,\orcidlink{0000-0002-1907-5964}} 
  \author{D.~M.~Asner\,\orcidlink{0000-0002-1586-5790}} 
  \author{T.~Aushev\,\orcidlink{0000-0002-6347-7055}} 
  \author{R.~Ayad\,\orcidlink{0000-0003-3466-9290}} 
  \author{V.~Babu\,\orcidlink{0000-0003-0419-6912}} 
  \author{Sw.~Banerjee\,\orcidlink{0000-0001-8852-2409}} 
  \author{M.~Bauer\,\orcidlink{0000-0002-0953-7387}} 
% ...
\collaboration{The Belle Collaboration}

\begin{abstract}
We present a measurement of the ratio of partial branching fractions of the semi-leptonic inclusive decays, $\BXulnu$ to $\BXclnu$, where $\ell = (e, \mu)$, using the full Belle sample of $772 \times 10^{6}$ $\BB$ pairs collected at the $\Y4S$ resonance. 
The identification of inclusive $\BXulnu$ decays is difficult due to the abundance of Cabibbo--Kobayashi--Maskawa-favored $\BXclnu$ events, which share a similar event topology. 
To minimize dependence on the modeling of these channels, a data-driven description of $\BXclnu$  is employed.
The ratio is measured via a two-dimensional fit to the squared four-momentum transfer to the lepton pair, and the charged lepton energy in the $B$ meson rest frame, where the latter must be larger than $1\gev$, covering approximately $86\%$ and $78\%$ of the $\BXulnu$ and $\BXclnu$ phase space, respectively. We find $\Delta \mathcal{B}(\BXulnu)/ \Delta \mathcal{B}(\BXclnu) = (1.99 \pm 0.17_{\rm stat} \pm 0.16_{\rm syst}) \times 10^{-2}$, where the uncertainties are statistical and systematic, respectively. We extract $|\Vub|/|\Vcb|$ using two theoretical calculations for the partial decay rate of $\BXulnu$, finding $(|\Vub|/|\Vcb|)^{\rm BLNP} = (9.81 \pm 0.42_{\rm stat} \pm 0.38_{\rm syst} \pm 0.51_{ \Delta\Gamma(\BXulnu)} \pm 0.20_{ \Delta\Gamma(\BXclnu)}) \times 10^{-2}$ and $
    (|\Vub|/|\Vcb|)^{\rm GGOU} = (10.06 \pm 0.43_{\rm stat} \pm 0.39_{\rm syst} \pm 0.23_{ \Delta\Gamma(\BXulnu)} \pm 0.20_{ \Delta\Gamma(\BXclnu)}) \times 10^{-2}$, where the third and fourth uncertainties are from the partial decay rates of $\BXulnu$ and $\BXclnu$, respectively.
In addition, we report the partial branching fractions  separately for charged and neutral $B$ meson decays, and for electron and muon decay channels. We place a limit on isospin breaking in $\BXulnu$ decays, and find no indication of lepton flavor universality violation in either the charmed or charmless mode. Furthermore, we unfold the $\BXulnu$ and $\BXclnu$ yields and report the differential ratio in lepton energy and four-momentum transfer squared. 
\end{abstract}

\noindent Belle Preprint 2023-17 \\ KEK Preprint 2023-30 \\ \\
\maketitle

\section{\label{sec:introduction}Introduction}
In the standard model (SM) of particle physics decays of quarks via the weak interaction are governed by the $3 \times 3$ unitary Cabibbo--Kobayashi--Maskawa (CKM) matrix \cite{Cabibbo:1963yz, Kobayashi:1973fv}.
The magnitudes of two of the elements, $|\Vub|$ and $|\Vcb|$, are extracted almost exclusively from measurements of semileptonic decays of $B$-mesons at the $B$-factories \cite{BaBar:2012thb, Belle:2013hlo, Belle:2010hep, BaBar:2010efp, BaBar:2007cke, BaBar:2007nwi, BaBar:2008zui, Belle:2010qug, Belle:2017rcc, Belle:2018ezy, CLEO:2002eqp, Belle:2005viu, BaBar:2016rxh, BaBar:2005acy, BaBar:2011xxm, Belle:2005uxj, Belle:2021eni, Belle:2006kgy, Belle:2006jtu, BaBar:2004bij, BaBar:2009zpz, Belle:2021idw} with LHCb measuring the ratio $|\Vub| / |\Vcb|$ via decays of the $B_s$ and $\Lambda_b$ hadrons~\cite{LHCb:2020ist, LHCb:2015eia}. As the hadronic and leptonic parts of the amplitudes can be factorized, theoretical predictions of the semileptonic decay rate are better understood than purely hadronic channels, which suffer from uncertainties arising from the strong interaction of the final state hadrons. The value of $|\Vub|$ is also accessible via purely leptonic decays, $B^- \to \mu^- \nub, B^- \to \tau^- \nub$; however, such channels are either helicity suppressed, or challenging to access experimentally due to the presence of multiple neutrinos
\cite{Belle:2019iji, Belle:2012egh, Belle:2015odw, BaBar:2009wmt, BaBar:2012nus}. 
Present measurements of $\mathcal{B}(B^- \to \tau^- \nub)$ allow for a determination of $|\Vub|$ with an uncertainty of $16\%$ \cite{FlavourLatticeAveragingGroupFLAG:2021npn}.

The CKM matrix is heavily over-constrained, allowing for powerful tests of the SM and constraints on beyond-SM physics via the comparison of quantities extracted from loop-level processes, or (semi)tauonic $B$ decays, which are expected to be sensitive to new-physics effects, and other tree-level processes, expected to be insensitive to new-physics effects. At present, such tests of the CKM sector are limited by the precision of the ratio $|\Vub| / |\Vcb|$ driven by the uncertainty of $|\Vub|$ determinations \cite{Charles:2004jd, UTfit:2006vpt}.

Measurements of $|V_{ub}|$ and $|V_{cb}|$ can be grouped into two experimentally and theoretically complementary strategies, known as exclusive, focusing on distinct final states, and inclusive, considering the sum of all possible final states. At present the world averages of these approaches show a $1.4\sigma$ and $2.5\sigma$ tension between inclusive and exclusive measurements in $|V_{ub}|$ and $|V_{cb}|$ \cite{ParticleDataGroup:2022pth}, respectively, with the inclusive measurements providing the larger values. Constraints from global fits of the CKM favor the inclusive $|V_{cb}|$ and exclusive $|V_{ub}|$ values \cite{Charles:2004jd}. The ratio $|V_{ub}|/|V_{cb}|$ determined separately with the exclusive and inclusive averages is in excellent agreement, and in reasonable agreement with the average of the direct determinations by LHCb \cite{LHCb:2020ist, LHCb:2015eia}. The current averages are \cite{ParticleDataGroup:2022pth, Charles:2004jd}:

{
\renewcommand{\arraystretch}{1.2}
\begin{tabular}{lr}
\\
$|V_{ub}|_{\rm exc.} = (3.70 \pm 0.10 \pm 0.12) \times 10^{-3}$,  & \cite{ParticleDataGroup:2022pth} \\
$|V_{ub}|_{\rm inc.} = (4.13\pm 0.12^{+0.13}_{-0.14} \pm 0.18) \times 10^{-3}$, & \cite{ParticleDataGroup:2022pth} \\
$|V_{ub}|_{\rm CKM} = (3.64^{+0.07}_{-0.07}) \times 10^{-3}$,  & \cite{Charles:2004jd}\nonumber\\
\\
$|V_{cb}|_{\rm exc.} = (39.4\pm 0.8) \times 10^{-3}$, & \cite{ParticleDataGroup:2022pth}\nonumber\\ 
$|V_{cb}|_{\rm inc.} = (42.2\pm 0.8) \times 10^{-3}$ , & \cite{ParticleDataGroup:2022pth}\nonumber\\
$|V_{cb}|_{\rm CKM} = (41.1^{+0.7}_{-0.4}) \times 10^{-3}$,  & \cite{Charles:2004jd} \nonumber\\
\\
$|V_{ub}|/|V_{cb}|_{\rm exc.} = (9.4 \pm 0.5) \times 10^{-2}$, & \cite{ParticleDataGroup:2022pth}\nonumber\\
$|V_{ub}|/|V_{cb}|_{\rm inc.} = (9.8 \pm 0.6) \times 10^{-2}$, & \cite{ParticleDataGroup:2022pth} \\
$|V_{ub}|/|V_{cb}|_{\rm LHCb} = (8.4 \pm 0.7) \times 10^{-2}$. & \cite{ParticleDataGroup:2022pth} \\ \\
\end{tabular}
}

\noindent Here, if a single uncertainty is provided it includes experimental and theoretical contributions, if two are provided these contributions are separated, and if three are provided they are experimental, theoretical, and given by the spread in theoretical determinations which will be further discussed below.

Exclusive measurements rely on form factor parameterizations of $\Bbar \to \pi \ell \nub$ and $\Bbar \to D^{(*)} \ell \nub$ decays \cite{Bourrely:2008za, Boyd:1994tt}, with inputs from lattice QCD at high four-momentum transfer squared, $q^2$ (see Ref.~\cite{FlavourLatticeAveragingGroupFLAG:2021npn} for a detailed review). These methods are mature and allow for determinations of $|\Vub|$ and $|\Vcb|$ with theoretical precision of $3\%$ and $1-2\%$, respectively, \cite{HFLAV:2019otj} \footnote{Note that in each case some of the uncertainty reported as experimental has theoretical origins.}.

Although inclusive determinations of both $|\Vub|$ and $|\Vcb|$ rely on an Operator Product Expansion (OPE) in the framework of the Heavy Quark Expansion \cite{Chay:1990da, Shifman:1986mx, Bigi:1992su}, treatment of both channels diverges due to experimental difficulties in measuring $\Bbar \to X_u \ell \nub$ decays. $|\Vcb|$ is commonly extracted from measurements of the partial branching fraction of $\Bbar \to \X_c \ell \nub$, and moments of the lepton momentum and hadronic mass spectra \cite{Belle:2008fsc, Gambino:2013rza, Bauer:2004ve, HFLAV:2019otj}, simultaneously extracting the $b$ quark mass $m_b$, and non-perturbative parameters of the OPE up to $\mathcal{O}(1/m_b^3)$. Recently a determination exploiting the reparametrization invariance of $q^2$ moments in order to reduce the number of free parameters at $\mathcal{O}(1/m_b^4)$ \cite{Fael:2019umf} has been performed~\cite{Bernlochner:2022ucr}. 

Inclusive determinations of $|\Vub|$ are complicated by the overwhelming background of $\Bbar \to X_c\ell\nub$ decays which have a similar event topology as the signal $\Bbar \to X_u\ell\nub$ channel, i.e.  a single high-energy lepton and a hadronic system. A clean sample of $\Bbar \to X_u\ell\nub$ events can only be selected by restricting the measurement to regions of phase space in which the charmed transition is kinematically forbidden. However, in such regions the local OPE does not converge well and non-perturbative shape functions need to be introduced \cite{Gambino:2020jvv, ParticleDataGroup:2020ssz, HFLAV:2019otj} to describe the Fermi motion of the $b$ quark within the $B$ meson. The leading shape function is expected to be universal for heavy-to-light transitions and can thus be constrained by measurements of the $B\to X_s\gamma$ photon energy spectrum. There are several models available, differing in their treatment of perturbative and non-perturbative parameters \cite{Gambino:2007rp, Lange:2005yw, DeFazio:1999ptt, Aglietti:2007ik, Andersen:2005mj}. While these methods individually predict partial rates with a precision of $6-10\%$ the spread between the methods is large and is taken by Ref.~\cite{ParticleDataGroup:2020ssz} as an additional uncertainty on the inclusive $|\Vub|$ world average, where it is dominant. In an effort to reduce these uncertainties measurements in Refs.~\cite{Belle:2021eni, Belle:2009pop, BaBar:2011xxm, BaBar:2016rxh} target the $\Bbar \to X_u \ell \nub$ spectra deep into the charm dominated region, with the most inclusive selections, covering approximately $86\%$ of the total rate \cite{DeFazio:1999ptt}. State-of-the-art measurements \cite{Belle:2021ymg} target differential spectra which allow for more model-independent determinations of $|\Vub|$ \cite{Gambino:2016fdy, Ligeti:2008ac, Bernlochner:2011di}. 

Hadronic tagging, in which the companion $B$ produced in the $e^+ e^- \to \Y4S \to \BB$ process is fully reconstructed in a hadronic mode, has proven to be a key experimental technique in inclusive $|\Vub|$ and $|\Vcb|$ determinations; enabling the measurement of properties of the hadronic $X$-system, and quantities associated with the neutrino, such as $q^2$. In this article we present the first direct measurement of $\Delta \mathcal{B}(\Bbar \to X_{u} \ell \nub) / \Delta \mathcal{B}(\Bbar \to X_{c} \ell \nub)$ performed on a hadronic tagged sample. By measuring the ratio directly we reduce experimental uncertainties originating from calibration of the tagging algorithms and lepton identification performance which can be sizable. Given the common dependence on $m_b$ and non-perturbative heavy-quark expansion parameters, we additionally expect that the ratio of partial branching fractions will allow for direct extraction of the CKM elements ratio, $|\Vub| / |\Vcb|$, with reduced theoretical uncertainty as well.

Several recent results have indicated poor agreement between data and $\Bbar \to X_c \ell \nub$ Monte Carlo (MC) modeling in key kinematic quantities,
 which may bias the extraction of the charmless semileptonic branching fraction \cite{Belle:2021idw, Bernlochner:2014dca, BaBar:2016rxh, Belle:2021eni}. In extracting $\Delta \mathcal{B}(\Bbar \to X_{u} \ell \nub)$ we therefore take a novel data-driven approach to constraining the $\Bbar \to X_c \ell \nub$ contribution.

The remainder of this article is organized as follows: section \ref{sec:exp} gives an overview of the experimental apparatus, data sample and simulated samples used in this analysis; section \ref{sec:analysis_strategy} outlines the event reconstruction and selection; section \ref{sec:measurement} details the measurement procedure with section \ref{sec:syst} describing the evaluation of systematic uncertainties. Finally, in sections \ref{sec:results} and \ref{sec:unfolding} the results are presented and discussed. 

Natural units, $\hbar = c = 1$ are used throughout this article. Inclusion of charge-conjugate mode decays is implied unless otherwise stated.

\section{\label{sec:exp} Experimental Apparatus and Data Samples}

We use the full \belle data sample of $711\invfb$ \cite{Belle:2012iwr} of integrated luminosity at the $\Y4S$ resonance, equivalent to ($772 \pm 10) \times 10^6 \BB$  pairs, produced by the KEKB accelerator complex \cite{Abe:2013kxa,Kurokawa:2001nw}. A further $89\invfb$ \cite{Belle:2012iwr} collected $60\mev$ below the $\Y4S$ resonance, hereafter referred to as off-resonance data, is used to study and derive correction factors for continuum processes $e^+ e^- \to \qqbar$, where $q = (u,d,s,c)$.

The Belle detector is a large-solid-angle magnetic spectrometer that consists of a silicon vertex detector (SVD), a 50-layer central drift chamber (CDC), an array of aerogel threshold Cherenkov counters (ACC), a barrel-like arrangement of time-of-flight (TOF) scintillation counters, and an electromagnetic calorimeter comprising CsI(Tl) crystals (ECL). All of these detectors are located inside a superconducting solenoid coil that provided a $1.5 \textrm{T}$ magnetic field. An iron flux return yoke located outside the coil is instrumented with resistive-plate chambers (KLM) to detect \KL mesons and to identify muons.
A more detailed description of the detector is provided in Ref.~\cite{Belle:2000cnh}.

Monte Carlo simulated samples of $\Y4S \to \BB$ and continuum processes are generated using the \texttt{EvtGen} generator \cite{Lange:2001uf} and corrected for electromagnetic final-state radiation by \texttt{PHOTOS} \cite{Barberio:1990ms}. The interactions of the particles with the detector are simulated using \texttt{Geant3}. \cite{Brun:1987ma}. The simulated samples correspond to approximately ten and six times the expected yield of $\BB$ and continuum events, respectively, in the \belle sample.

Semileptonic $\BXulnu$ and $\BXclnu$ decays are modeled following the approach detailed in Ref.~\cite{Belle:2021ymg}. Charmless decays are modeled as a mixture of resonant and non-resonant contributions combined via a hybrid approach as proposed by Ref.~\cite{Ramirez:1989yk}. The channels are normalized to the world averages from Ref.~\cite{ParticleDataGroup:2020ssz}. The resonant decays $\Bbar \to \pi \ell \nub$, $\Bbar \to \rho \ell \nub$, $\Bbar \to \omega \ell \nub$ are modeled via the expansion of Bourrely, Caprini, and Lellouch (BCL) \cite{Bourrely:2008za}. For $\Bbar \to \pi \ell \nub$ we adopt the form factor central values and uncertainties from the global fit of Ref.~\cite{FermilabLattice:2015mwy}. For $\Bbar \to \rho \ell \nub$ and $\Bbar \to \omega \ell \nub$ decays we adopt the values of Ref.~\cite{Bharucha:2015bzk}. The decays $\Bbar \to \eta \ell \nub$ and $\Bbar \to \eta^\prime \ell \nub$ are modeled via the light-cone sum rule predictions of Ref.~\cite{Duplancic:2015zna}. 

Non-resonant $\BXulnu$ decays, hereafter referred to as $\Bbar \to x_u \ell \nub$, are simulated using the  model of DeFazio and Neubert (DFN) \cite{DeFazio:1999ptt}. The triple differential rate of this model is a function of the four-momentum transfer squared, the lepton energy in the $B$ rest-frame ($E_\ell^B$), and the hadronic invariant mass ($M_X$) of the $X_u$ system at next-to-leading order precision in the strong coupling constant $\alpha_s$. The rate is convolved with a non-perturbative shape function using an ad-hoc exponential model. The free parameters of the model are the $b$ quark mass in the Kagan-Neubert scheme \cite{Kagan:1998ym}, $m^{\rm KN}_b = (4.66 \pm 0.04) \gev$ and a non-perturbative parameter $a^{\rm KN}= 1.3 \pm 0.5$. The values of these parameters were determined in Ref.~\cite{Buchmuller:2005zv} from a fit to $\BXclnu$ and $\Bbar \to X_s \gamma$ decay properties. At leading order, the non-perturbative parameter $a^{\rm KN}$ is related to the average momentum squared of the $b$ quark inside the $\B$ meson and determines the second moment of the shape function. It is defined as $a^{\rm KN} = -3\overline{\Lambda}^2/\lambda_1 - 1$ with the binding energy $\overline{\Lambda} = m_B - m^{\rm KN}_b$, where $m_B$ is the $B$ meson mass, and the kinetic energy parameter $\lambda_1$. The hadronization of the parton-level $\BXulnu$ simulation is carried out using the JETSET algorithm \cite{Sjostrand:2000wi} to at least two final state mesons.
The resonant and non-resonant contributions are combined such that the sum of exclusive ($\Delta \mathcal{B}^{\textrm{exc}}_{ijk}$) and inclusive ($\Delta \mathcal{B}^{\textrm{inc}}_{ijk}$) contributions reproduce the inclusive predictions in a three-dimensional binning of the triple differential decay rate. For each bin inclusive events are assigned a weight, $w_{ijk}$, defined to be 
\begin{equation}
    w_{ijk} = \frac{\Delta \mathcal{B}^{\textrm{inc}}_{ijk} - \Delta \mathcal{B}^{\textrm{exc}}_{ijk}}{\Delta \mathcal{B}^{\textrm{inc}}_{ijk}}, \label{eq:hybrid}
\end{equation}
where $i,j,k$ run over the bins in $q^2, E_\ell^B, M_X$, defined by the bin boundaries:

\begin{align*}
    q^2 &: [0, 2.5, 5, 7.5, 10, 12.5, 15, 20, 25] \gevv, \\
    E_\ell^B &: [0, 0.5, 1, 1.25, 1.5, 1.75, 2, 2.25, 3] \gev, {\rm~and} \\
    M_X &: [0, 1.4, 1.6, 1.8, 2, 2.5, 3, 3.5] \gev. \\
\end{align*}
The branching fractions used in the simulation of the $\BXulnu$ channels are summarized in Table~\ref{tab:x_bf}.

\begin{table}[]
    \centering
    \caption{Simulated branching fractions for $\BXulnu$  and $\BXclnu$ decays. Non-resonant charmless decays are denoted as $\Bbar \to x_{u} \ell \nub$.}
    \label{tab:x_bf}
\begin{ruledtabular}
\begin{tabular}{l c c}
    Decay Channel & $\B^+$ [$\times 10^{-3}$] & $B^0$ [$\times 10^{-3}$] \\ \hline
    $\Bbar \to X_{u} \ell \nub$ & $2.21 \pm 0.31$ & $2.05 \pm 0.29$  \\ \hline
    \qquad $\Bbar \to \pi \ell \nub$ & $0.078 \pm 0.003$ & $0.150 \pm 0.006$ \\ 
    \qquad $\Bbar \to \rho \ell \nub$ & $0.158 \pm 0.011$ & $0.294 \pm 0.021$ \\ 
    \qquad $\Bbar \to \omega \ell \nub$ & $0.119 \pm 0.009$ & - \\ 
    \qquad $\Bbar \to \eta \ell \nub$ & $0.039 \pm 0.005$ & - \\ 
    \qquad $\Bbar \to \eta^\prime \ell \nub$ & $0.023 \pm 0.008$ & - \\ 
    \qquad $\Bbar \to x_{u} \ell \nub$ & $1.79 \pm 0.32$ & $1.60 \pm 0.30$   \\ \hline
    
    $\Bbar \to X_{c} \ell \nub$ & $108 \pm 4$ & $101 \pm 4$  \\ \hline
    \qquad $\Bbar \to D \ell \nub$ & $23.5 \pm 1\hphantom{0}$ & $23.1 \pm 1\hphantom{0}$ \\ 
    \qquad $\Bbar \to D^* \ell \nub$ & $56.6 \pm 2\hphantom{0}$ & $50.5 \pm 1\hphantom{0}$ \\ \hline

    \qquad  $\Bbar \to D_0 (\to D \pi) \ell \nub$ & $4.2 \pm 0.8$ & $3.9\pm 0.7$  \\ 
    \qquad  $\Bbar \to D_1^\prime (\to D^* \pi) \ell \nub$ & $4.2 \pm 0.8$ & $3.9\pm 0.8$   \\ 
    \qquad  $\Bbar \to D_1 (\to D^* \pi) \ell \nub$ & $4.2 \pm 0.3$ & $3.9\pm 0.3$  \\ 
    \qquad  $\Bbar \to D_1 (\to D \pi \pi) \ell \nub$ & $2.4\pm 1.0$  & $2.3\pm0.9$ \\  
    \qquad  $\Bbar \to D_2 (\to D^* \pi) \ell \nub$ & $1.2 \pm 0.1$ & $1.1\pm 0.1$ \\ 
    \qquad  $\Bbar \to D_2 (\to D \pi) \ell \nub$ & $1.8 \pm 0.2$ & $1.7\pm 0.2$  \\ \hline 
    
    \qquad $\Bbar \to  D \pi \pi \ell \nub$ & $0.6 \pm 0.6$  & $0.6 \pm 0.6$ \\ 
    \qquad $\Bbar \to  D^* \pi \pi \ell \nub$ & $2.2 \pm  1.0$  & $2.0 \pm 1.0$ \\ 
    \qquad $\Bbar \to  D \eta \ell \nub$ & $3.6\pm 2.0$  & $4.0 \pm 2.0$ \\ 
    \qquad $\Bbar \to  D^* \eta \ell \nub$ & $3.6\pm 2.0$  & $4.0 \pm 2.0$ \\
\end{tabular}
\end{ruledtabular}
\end{table}

The $\BXclnu$ rate is dominated by $\Bbar \to D \ell \nub$ and $\Bbar \to D^* \ell \nub$ decays. The $\Bbar \to D \ell \nub$ decays are modeled using the parameterization of Boyd, Grinstein, and Lebed (BGL) \cite{Boyd:1994tt} with form factor central values and uncertainties taken from the fit in Ref.~\cite{Belle:2015pkj}. For $\Bbar \to D^* \ell \nub$ decays we use the BGL implementation proposed by Refs \cite{Bigi:2017njr, Grinstein:2017nlq} with form factor central values and uncertainties from the fit to the measurement of Ref.~\cite{Ferlewicz:2020lxm}. Both channels are normalized to the average branching fraction of Ref.~\cite{HFLAV:2019otj}. Semileptonic decays to the four orbitally excited charmed mesons ($D_0^*, D_1^*, D_1, D_2^*$), hereafter collectively denoted as $D^{**}$, are modeled using the heavy-quark-symmetry-based form factors proposed in Ref.~\cite{Bernlochner:2016bci}. We simulate all $D^{**}$ decays using masses and widths from Ref.~\cite{ParticleDataGroup:2020ssz}. We adopt the branching fractions of Ref.~\cite{HFLAV:2019otj} and correct them to account for missing isospin-conjugated and other established decay modes, following the prescription of Ref.~\cite{Bernlochner:2016bci}. As the measurements were carried out in the $D^{**0} \to D^{(*)+} \pi^-$ decay modes we account for the missing isospin modes with a factor of 
\begin{equation}
    f_\pi = \frac{\mathcal{B}(D^{**0} \to D^{(*)+} \pi^-)}{\mathcal{B}(D^{**0} \to D^{(*)} \pi)} = \frac{2}{3}.
\end{equation}
The measurements of the $\mathcal{B}(\Bbar \to D_2^* \ell \nub)$ in Ref.~\cite{HFLAV:2019otj} are converted to only account for the $D_2^{*0} \to D^{*+} \pi^{-}$ decay. To also account for $D_2^{*0} \to D^{+} \pi^{-}$ contributions we apply a factor of \cite{ParticleDataGroup:2020ssz}
\begin{equation}
    f_{D_2^*} = \frac{\mathcal{B}(D_2^{*0} \to D^{*+} \pi^{-})}{\mathcal{B}(D_2^{*0} \to D^{+} \pi^{-})} = 1.54 \pm 0.15. 
\end{equation}
The world average of $\mathcal{B}(\Bbar \to D_1^* \ell \nub)$ given in Ref.~\cite{HFLAV:2019otj} combines measurements that show poor agreement, and the resulting p-value of the combination is below $0.01\%$. Notably, the measurement of Ref.~\cite{Belle:2007uwr} is in conflict with the measured branching fractions of Refs.~\cite{BaBar:2008dar, DELPHI:2005mot} and with the expectation of $\mathcal{B}(\Bbar \to D_1^* \ell \nub)$ being of similar size than $\mathcal{B}(\Bbar \to D_0 \ell \nub)$ \cite{Leibovich:1997em, Bigi:2007qp}. We perform our own average excluding Ref.~\cite{Belle:2007uwr} and use
\begin{equation}
    \mathcal{B}(B^- \to D_1^{*0}  (\to D^{*+} \pi^-) \ell \nub ) = (0.28 \pm 0.06) \times 10^{-2}.
\end{equation}
The world average of $\mathcal{B}(\Bbar \to D_1 \ell \nub)$ does not include contributions from prompt three-body decays of $D_1 \to D \pi \pi$. We account for these using a factor \cite{LHCb:2011poy}
\begin{equation}
    f_{D_1} = \frac{\mathcal{B}(D_1^0 \to D^{*+} \pi^{-})}{\mathcal{B}(D_1^0 \to D^{0} \pi^+ \pi^{-})} = 2.32 \pm 0.54.
\end{equation}

We subtract the contribution of $\mathcal{B}(\Bbar \to D_1( \to D\pi\pi) \ell \nub)$ from the measured non-resonant plus resonant $\mathcal{B}(\Bbar \to D \pi \pi \ell \nub)$ of Ref.~\cite{BaBar:2015zkb}. To account for missing isospin-conjugated modes of the three-hadron final states we adopt the prescription from Ref.~\cite{BaBar:2015zkb}, 
\begin{equation}
    f_{\pi\pi} = \frac{\mathcal{B}(D^{**0} \to D^{(*)0}\pi^+\pi^-)}{\mathcal{B}(D^{**0} \to D^{(*)}\pi\pi)} = \frac{1}{2} \pm \frac{1}{6}.
\end{equation}
The uncertainty takes into account the full spread of final states ($f_{0}(500) \to \pi \pi$ or $\rho \to \pi \pi$ result in $f_{\pi\pi} = 2/3 \textrm{ and } 1/3$ respectively) and the non-resonant three-body decays ($f_{\pi\pi} = 3/7$). We further assume that 

\begin{align}
    \mathcal{B}(D^*_2 \to D\pi) + \mathcal{B}(D^*_2 \to D^*\pi) &= 1, \nonumber \\
    \mathcal{B}(D_1 \to D^*\pi) + \mathcal{B}(D_1 \to D\pi\pi) &= 1, \nonumber \\
    \mathcal{B}(D_1^* \to D^*\pi) &= 1,  \nonumber \\
    \mathcal{B}(D_0 \to D\pi) &= 1.
\end{align}

For the remaining $\mathcal{B}(\Bbar \to D^{(*)} \pi\pi\ell\nub)$ contributions we use the measured value of Ref.~\cite{BaBar:2015zkb}. The remaining ``gap" between the sum of all considered exclusive modes and the inclusive $\BXclnu$ branching fraction ($\approx 0.8 \times 10^{-2}$ or $7-8\%$ of the total $\BXclnu$ branching fraction) is filled in equal parts with $\Bbar \to D \eta \ell \nub$ and $\Bbar \to D^* \eta \ell \nub$ decays where we take the error to be uniform between zero and twice the nominal branching fraction. We simulate $\Bbar \to D^{(*)} \pi \pi \ell \nub$ and  $\Bbar \to D^{(*)} \eta \ell \nub$ final states assuming that they are produced by the decay of two broad resonant states $D^{**}_{\textrm{Gap}}$ with masses and widths identical to $D_1^*$ and $D_0$. Although there is currently no experimental evidence for decays of charm $1P$ states into these final states or the existence of such an additional broad state (e.g. a $2S$) in semileptonic transitions, this description provides a better kinematic distribution of the initial three-body decay $\Bbar \to D_{\textrm{Gap}}^{**} \ell \nub$, than e.g. a model based on the equi-distribution of all final-state particles in phase space. For the form factor of the $D_{\textrm{Gap}}^{**}$ modes we adopt the same description as for the $D^{**}$ modes \cite{Bernlochner:2016bci}. We neglect the small contribution from $\Bbar \to D_s^{(*)} K \ell \nub$ which has a branching fraction of $(5.9 \pm 1.0)\times 10^{-4}$ \cite{ParticleDataGroup:2020ssz, BaBar:2010ner, Belle:2012ccr}. The used $\BXclnu$ branching fractions are summarized in Table \ref{tab:x_bf}.

\section{\label{sec:analysis_strategy} Event Reconstruction}

The analysis is performed in the \belletwo\ analysis software framework \cite{Kuhr:2018lps, collaborationBelleIIAnalysis2021} through the use of the \belle to \belletwo\ conversion software package \cite{Gelb:2018agf}. 

\subsection{Tag Side Reconstruction}

One $B$ meson in the event is reconstructed with the hadronic Full Event Interpretation (FEI) algorithm detailed in Ref.~\cite{Keck:2018lcd}. The FEI algorithm applies a six step hierarchical reconstruction process beginning by identifying all tracks and clusters in the event as final state particles ($e, \mu, K, \pi, \gamma$), then consecutively reconstructing heavier intermediate stage mesons ($\pi^0, \KS, \jpsi, D_{(s)}, D^{*}_{(s)}$) before finally reconstructing $B$ candidates. In each stage of the reconstruction a loose preselection is applied to keep computing time reasonable. A fast vertex fit is then performed before a gradient boosted decision tree classifier estimates the signal probability for each candidate \cite{Keck:2017gsv}. Finally this signal probability is used to select $\mathcal{O}(10)$ candidates that proceed to the next stage. In total $\mathcal{O}(10000)$ channels are reconstructed.

To select well reconstructed tags we require that the candidates have a beam-energy constrained mass of 
\begin{equation}
    M_{\textrm{bc}} = \sqrt{E^2_{\textrm{beam}} - |\vec{p}^{\,*}_{B_{\textrm{tag}}}|^2} > 5.27 \gev,
\end{equation}
where $E_\textrm{beam} = \sqrt{s}/2$ denotes half the center-of-mass (c.m.) energy and $\vec{p}^{\,*}_{B_{\textrm{tag}}}$ is the three momentum of the $B_{\textrm{tag}}$ candidate in the c.m. frame. This condition is relaxed to $M_{\textrm{bc}} > 5.24 \gev$ when analyzing off-resonance data. We further consider the energy difference from nominal, $\Delta E$, given by
\begin{equation}
    \Delta E = E^*_{B_{\textrm{tag}}} - E_{\textrm{beam}},
\end{equation}
where $E^*_{B_{\textrm{tag}}}$ is the energy of the $B_{\textrm{tag}}$ in the c.m. frame, and we impose the requirement, $-0.1 < \Delta E < 0.15 \gev$. Finally, each $B_{\textrm{tag}}$ candidate is associated with a final classifier score from the FEI, $C_{\textrm{FEI}}$. We require that $C_{\textrm{FEI}} > 0.01$. All tracks and clusters not used in reconstructing the tag are assigned to the signal side. 

\subsection{Signal Side Reconstruction}

The signal side of the event is defined by the presence of a well identified lepton. Leptons are selected by use of a likelihood ratio $\mathcal{L}_{\textrm{LID}}$ that combines information from multiple sub-detectors. Electrons are primarily identified by the ratio of energy deposition in the ECL to the magnitude of momentum of the reconstructed track, the energy loss in the CDC, the shower shape in the ECL, the quality of the geometric matching of the track to the shower position in the ECL, and the photon yield in the ACC  \cite{Hanagaki:2001fz}. Muons are identified from charged track trajectories extrapolated to the KLM sub-detector. The key features are the difference between expected and measured penetration depth and the transverse deviation of KLM hits from the extrapolated trajectory \cite{Abashian:2002bd}.  To ensure decay channel orthogonality, we require electron candidates to fail the muon selection criteria. Electrons and muons are identified with efficiencies of about $90\%$ and pion-to-lepton misidentification (``fake") rates of $0.25\%$ \cite{Hanagaki:2001fz} and $1.5\%$ \cite{Abashian:2002bd}, respectively. For electrons we correct for bremsstrahlung by 
searching for photon candidates with an energy less than $1 \gev$ within a cone of $5^\circ$ around the initial momentum direction of the electron track. The candidate with the highest energy is taken as radiative and its momentum re-summed to the electron candidate.

We require that lepton candidates have momenta in the c.m. frame exceeding $0.5 \gev$, and pass through the barrel of the detector, corresponding to an angular acceptance of $\theta_{\textrm{lab}} \in (35^\circ, 125^\circ)$ and $\theta_{\textrm{lab}} \in (25^\circ, 145^\circ)$ for electrons and muons, respectively, where $\theta_{\textrm{lab}}$ is the polar angle of the lepton candidate with respect to the direction opposite to the positron beam.

The remaining tracks and neutral energy depositions in the event are inclusively summed to form the $X$ system. To improve the resolution of quantities associated with the $X$ system we impose selection criteria on both tracks and photons, where photons are reconstructed from neutral energy depositions in the ECL that have not been matched to a track.
To veto beam-background-induced photons we discard photons that do not meet minimum energies of $100,50,150 \mev$ when found in the forward end-cap ($12.4^\circ \leq \theta < 31.4^\circ$), barrel ($32.2^\circ \leq \theta < 128.7^\circ$), and backward end-cap ($130.7^\circ \leq \theta < 155.1^\circ$) of the ECL, respectively.

For tracks, we require $|\vec{p}^{\,*}| < 3.2 \gev$, where $\vec{p}^{\,*}$ is the c.m. frame three-momentum, and a transverse momentum, $p_t$, dependent selection on the distance between the interaction point and point of closest approach of each track to the $z$-axis, defined to be opposite to the positron beam direction, $|dz|$, and in the plane transverse $z$-axis, $|dr|$. This ensures that tracks originate near the interaction point. The selections are:

\begin{align*}
    p_t < 0.25 \gev &: |dr| < 20\cm, |dz| < 100\cm, \\
    0.25 \leq p_t < 0.50 \gev &: |dr| < 15\cm, |dz| < 50\cm, \\
     p_t \geq 0.50 \gev &: |dr| < 10\cm, |dz| < 20\cm.
\end{align*}

Low momentum tracks can curl within the tracking detectors and be reconstructed as multiple tracks causing non-zero net event charges. We therefore train a gradient boosted decision tree \cite{Keck:2017gsv} to compare pairs of tracks and identify duplicates. The variables considered are error-weighted differences of the five track helix parameters, the product of the track charges, the angle between the tracks, and error-weighted differences in the transverse and longitudinal momenta. If multiple tracks are identified to have originated from the same particle they are ranked by $25|dr|^2 + |dz|^2$ with all but the lowest value track removed. The classifier is trained separately for the early and late SVD geometry \cite{Natkaniec:2006rv}. 

We attempt to identify kaons in the rest of the event to tag $b \to c \to s$ cascade decays. Kaons are identified with use of a likelihood ratio, 
where the dominant features contributing to the kaon and pion likelihoods are the energy loss in the CDC and either the time of flight information in the TOF for low-$p_t$ tracks, or the Cherenkov light recorded in the ACC for high-$p_t$ tracks. They are identified with an efficiency of $88\%$ at a pion to kaon fake rate of $8.5\%$ \cite{Nakano:2002jw}.  

Candidate $\KS$-mesons  are reconstructed from pairs of charged pions and selected by use of a multivariate NeuroBayes classifier \cite{Feindt:2006pm, Belle:2018xst}. We make no attempt to reconstruct $\KL$ particles.
The four-momentum of the $X$ system is calculated as
\begin{equation}
    p_X = \sum_{i \in \{\textrm{tracks, photons}\}} (\sqrt{m_{i}^2 + |\vec{p}_{i}|^2}, \vec{p}_{i}) 
\end{equation}
where  $m_{i}$ is the nominal mass of the particle hypothesis assigned to track $i$ or $0$ for photons. The four-momentum of the neutrino is inferred to be the missing four-vector in the event, defined as:
\begin{equation}
    p_\textrm{Miss} = p_{e^+e^-} - p_{B_{\rm tag}} - p_\ell - p_X.
\end{equation}
where $p_{e^+e^-}$, $p_{B_{\rm tag}}$, $p_\ell$, are the four-vectors of the c.m., $\B_{\rm tag}$, and lepton, respectively. We additionally calculate the momentum transfer squared as
\begin{equation}
    q^2 = ((|\vec{p}_{\textrm{Miss}}|, \vec{p}_{\textrm{Miss}}) +  p_\ell)^2,
\end{equation}
where $\vec{p}_\textrm{Miss}$ is the missing momentum three-vector. This definition of $q^2$ is found to have a resolution $16\%$ smaller than a definition which considers also the missing energy. 

\subsection{Event Selection}

We further select events by requiring that the sum of the charges of the tag, lepton, and tracks associated to the $X$ system is zero: consistent with a well reconstructed $\Y4S \to B_{\textrm{tag}} B (\to X \ell \nub)$ event. To suppress candidate events with a fake lepton or events where we have identified a lepton from a secondary $c \to s \ell \nub$ decay we require that the charge of the lepton candidate is consistent with a prompt semileptonic decay of $B_{\textrm{sig}}$, where the flavor of $B_{\textrm{sig}}$ has been inferred from that of the $B_{\rm tag}$. We make no allowance for $B^0$-mixing.

If after the enforcement of these selection criteria multiple candidates are present we select the candidate with the highest $C_{\textrm{FEI}}$. If any tracks assigned to the $X$ system pass the electron or muon likelihood requirements imposed on the signal lepton, the event is vetoed, as we cannot unambiguously ascertain which lepton originates from the primary $B$ decay and which originates from a secondary decay. Additionally, the presence of a secondary lepton indicates the likely presence of a secondary neutrino. The momentum carried away by such a neutrino would be incorrectly assigned to the neutrino produced in the primary semileptonic decay, worsening the resolution of quantities calculated from $p_\textrm{Miss}$. We further require that the number of charged kaons in the event is consistent with $|N_{K^+} - N_{K^-}| \leq 1$, allowing for a single kaon produced via a $b \to c \to s$ cascade decay and additional pairs produced via pair production in what is commonly referred to as $\ssbar$ popping.

We veto continuum events by means of boosted decision tree multivariate classifiers \cite{Keck:2017gsv} trained on $30$ variables describing the event shape. As $\BB$ pairs are produced nearly at rest in the c.m. frame their decays are expected to proceed spherically. Lighter quark pairs are produced with a significant boost leading to jet-like decays.
The variables under consideration are: CLEO cones \cite{CLEO:1995rok}, $B$ meson thrust angles and magnitudes, and modified Fox-Wolfram moments \cite{Belle:2003fgr}. Four classifiers are trained, two each for the early SVD and late SVD geometry \cite{Natkaniec:2006rv}. The classifiers are cross-applied on simulated samples. For data, one of the two SVD geometry appropriate classifiers is randomly selected. The selection rejects $92.9\%$ of simulated continuum events and retains $94.0\%$ and $89.0\%$ of simulated $\BXulnu$ and $\BXclnu$ events, respectively.

We restrict the allowed lepton energy in the signal $B$ meson rest frame to the region of interest, $\lepp > 1 \gev$. Here we define $\lepp = |\vec{p}^{\, B_{\rm sig}}_{\ell}|$, neglecting the small contribution of the lepton mass to the energy, and inferring the signal $B$ momentum from the tag $B$. In the c.m. frame this is given by
\begin{equation}    
    \vec{p}^{\,*}_{B_{\textrm{sig}}} = - \vec{p}^{\,*}_{B_{\textrm{tag}}}. 
\end{equation}

We further purify the sample used for the $\BXulnu$ yield extraction by suppressing $\Bzb \to D^{*+}( \to D^0 \pi^+) \ell \nub$ events via a slow-pion veto. We search for charged pions with momentum between $50$ and $200 \mev$ in the c.m. frame. If such a pion is found, we inclusively reconstruct the $D^*$ meson. Due to the small mass difference of the $D^{*+}$ and $D^0$ mesons, the pion is expected to be almost co-linear in the lab frame with the $D^{*+}$. We assume the $D^*$ momentum to lie along the flight direction of the slow pion and estimate its energy to be
\begin{equation}
    E_{D^*} = \frac{m_{D^*}}{m_{D^*} - m_{D^0}} E_{\pi}.
\end{equation}
We then calculate the missing mass squared from the estimated $D^*$ four vector,
\begin{equation}
    M^2_{\textrm{Miss}}(D^*) = (p_{e^+e^-} - p_{B_{\rm tag}} - p_{\ell} - p_{D^*})^2.    
\end{equation}
If the slow pion originates from a $D^*$ decay $M^2_{\textrm{Miss}}(D^*)$ is expected to be consistent with zero. If the slow pion does not originate from the decay of a $D^*$ the incorrectly estimated $D^*$ four vector leads to negatively skewed $M^2_{\textrm{Miss}}(D^*)$ values. We veto events with $M^2_{\textrm{Miss}}(D^*) > -2 \gevv$.

Finally, we consider the missing mass squared of the event reconstructed using the full $X$ system defined as
\begin{equation}
    M^2_{\textrm{Miss}} = (p_{e^+e^-} - p_{B_{\textrm{tag,cons.}}} - p_{\ell} - p_{X})^2,    
\end{equation}
where we constrain the energy of the $B_{\textrm{tag}}$ in the c.m. frame to half the c.m. energy. For prompt semileptonic events $M^2_{\textrm{Miss}}$ should be consistent with zero. For $\BXclnu$ events and other backgrounds in the extraction of $\BXulnu$ a skew towards positive values is observed. We impose the requirement $|M^2_{\textrm{Miss}}| < 0.43 \gevv$, which has been optimized to minimize the total experimental uncertainty on the measurement of the ratio of partial branching fractions for $\lepp > 1.0 \gev$.

Particle identification efficiencies have been investigated in dedicated control mode studies:
we correct the identification efficiency in MC for $e^+, \mu^+, K^+, \KS, \pi^{\pm}_{\textrm{slow}}$ particles and the $\pi^+, K^+ \to e^+, \mu^+, (K^+)$ fake rates.

To correct for differences in performance of the continuum suppression classifier on data and MC and correct potential overall normalization issues we study the off-resonance data sample. The data to MC disagreement is observed to be lepton flavor and momentum dependent. We fit linear calibration functions to the ratio of data to MC yields in bins of $\lepp$ for the electron and muon modes separately and apply correction factors to on-resonance continuum MC events.

The reconstruction efficiency of the FEI algorithm is known to vary between data and MC \cite{Keck:2018lcd}. To account for this, we calibrate in situ to the $\Bbar \to X \ell \nub$ yield. For each of the 29 (26) reconstructed charged (neutral) $B$ channels under consideration we perform a binned extended likelihood fit to the $M_{\textrm{bc}}$ spectrum using histogram PDFs. The fit considers three components: well reconstructed $B$ decays, badly reconstructed $B$ decays, and the continuum yield which is fixed. Candidates are considered well reconstructed if at most one photon has been incorrectly assigned. For each channel two correction factors are derived by comparing the fitted yield of the \BB components with the number reconstructed in MC. To perform the calibration, we broaden our event selection to $M_{\textrm{bc}} > 5.24 \gev$, $\lepp > 0.7 \gev$, and consider only tag candidates with the highest $C_{\textrm{FEI}}$ after a loose preselection. We measure mean correction factors of   $0.68 \pm 0.001_{\rm stat} \pm 0.023_{\rm syst}$ for well reconstructed $B_{\rm tag}$ and $0.87 \pm 0.002_{\rm stat} \pm 0.041_{\rm syst}$ for poorly reconstructed $B_{\rm tag}$. The reconstruction efficiency of the simulated $\BXulnu$ samples by component is given in Table~\ref{tab:xulnu_sel_composition}. 
\begin{table}[]
    \centering
    \caption{Reconstruction efficiency of the simulated $\BXulnu$ sample by component. Approximately $68\%$ of reconstructed $\BXulnu$ events are expected to originate from the decay of charged $B$ mesons. Uncertainties are statistical.}
    \label{tab:xulnu_sel_composition}
\begin{ruledtabular}
\begin{tabular}{l c c}
    Decay Channel & $\B^+$ [$\%$] & $B^0$ [$\%$] \\ \hline
    $\Bbar \to \pi \ell \nub$ & $0.140(3)$ &   $0.076(2)$ \\
    $\Bbar \to \rho \ell \nub$ & $0.138(2)$ &  $0.077(1)$ \\
    $\Bbar \to \eta \ell \nub$ & $0.114(3)$ & -  \\
    $\Bbar \to \omega \ell \nub$ & $0.120(2)$ & -  \\
    $\Bbar \to \eta^\prime \ell \nub$ &  $0.085(4) $ &  -  \\
    $\Bbar \to x_u \ell \nub$ &  $0.0865(4)$ & $0.0440(3)$ \\
\end{tabular}
\end{ruledtabular}

\end{table}

\begin{figure*}[htb]

\includegraphics[width=0.45\textwidth]{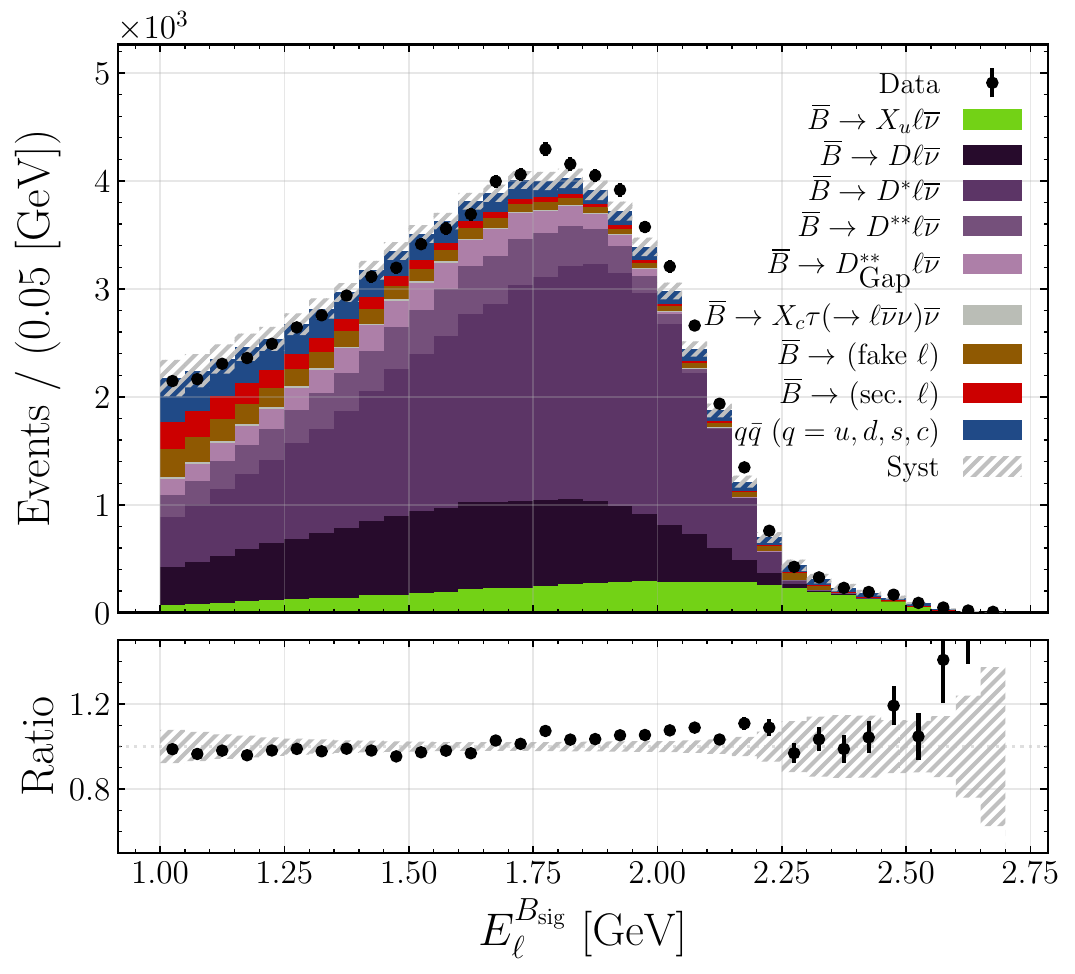}%
\includegraphics[width=0.45\textwidth]{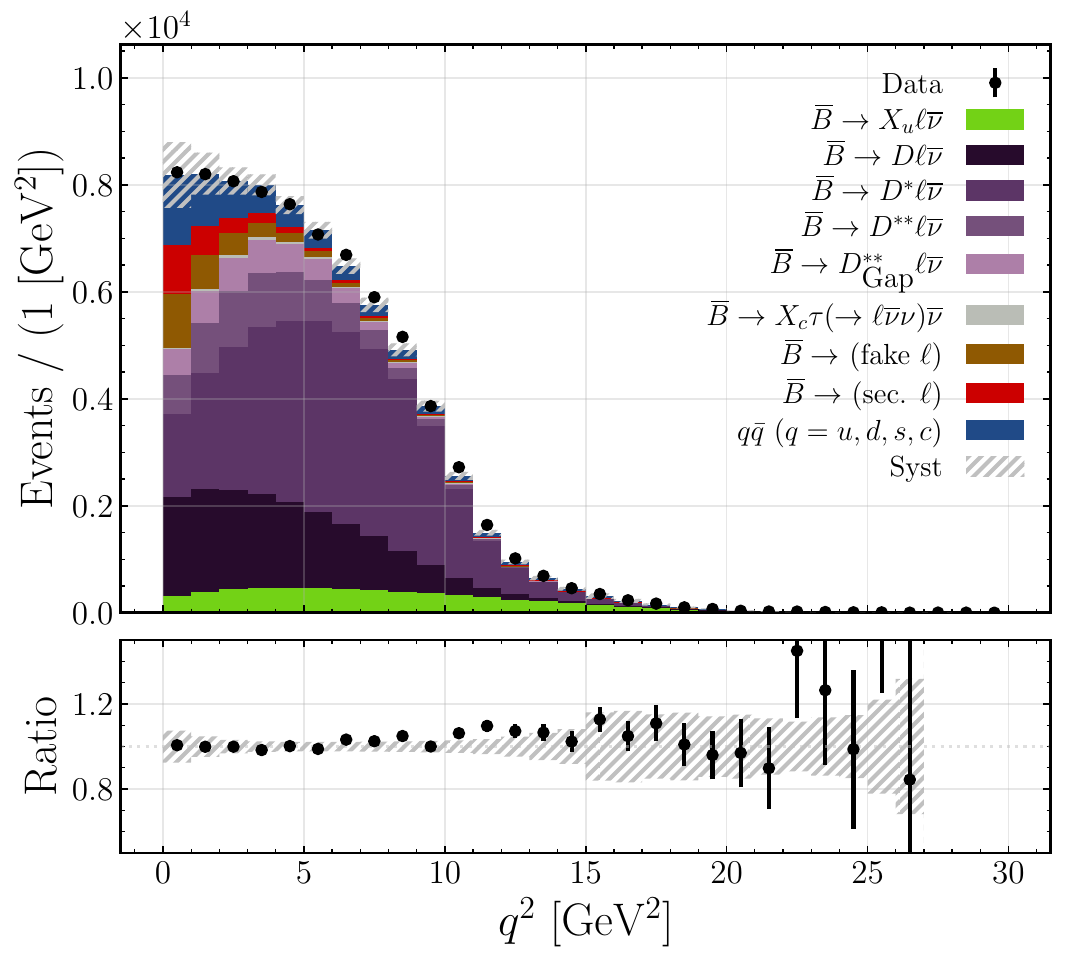}  \\
\includegraphics[width=0.45\textwidth]{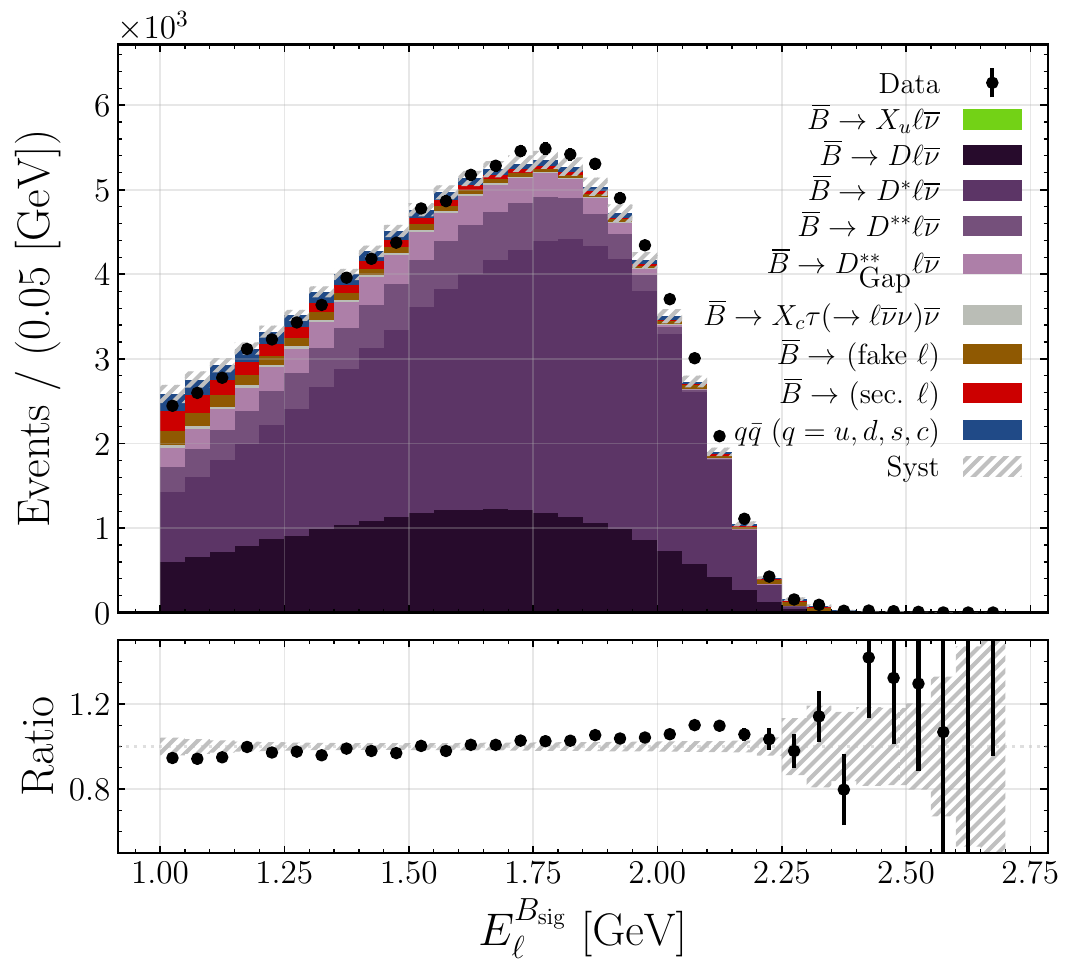} 
\includegraphics[width=0.45\textwidth]{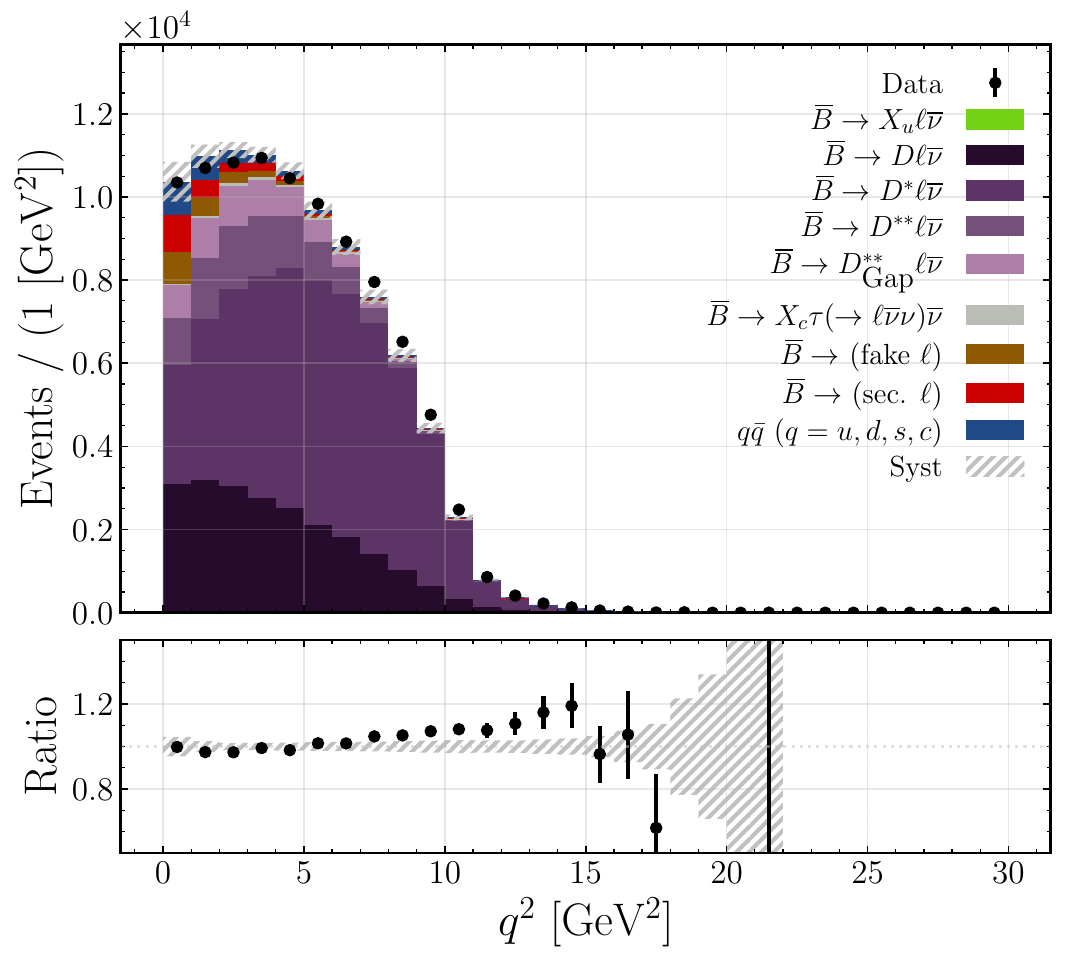} \\

\caption{\label{fig:data_mc_agreement} Reconstructed $\lepp$ and $q^2$ spectra for the $\BXulnu$ enhanced (\textit{top}) and depleted (\textit{bottom}) sub-samples. The error bands of simulated samples incorporate the full set of systematic uncertainties discussed in section \ref{sec:syst}.}
\end{figure*}

After event selection we retain $172312$ events. We further separate the sample into $\BXulnu$ enhanced and depleted sub-samples based on the number of kaons found in the $X$ system. The latter is used to derive a data-driven description of the remaining $\BXclnu$ contamination in the $\BXulnu$ enhanced sample. To allow for potential $\ssbar$ popping we assign events with even values of $N_{K^\pm} + N_{\KS}$ to the $\BXulnu$ enhanced sub-sample ($76835$ events) while remaining events are assigned to the depleted sub-sample. This splitting criterion identifies $\BXulnu$ ($\BXclnu$) decays with $93\%$ ($40\%$) efficiency in simulation. 
The majority of the continuum events and other backgrounds fall into the $\BXulnu$ enhanced sample. The data-MC agreement of the selected candidates in kinematic variables of interest, $\lepp, q^2$ is shown in Fig. \ref{fig:data_mc_agreement}. The agreement in the $\BXulnu$ enhanced sample is well replicated in the $\BXulnu$ depleted sample.

\section{\label{sec:measurement} Measurement Procedure}

\begin{figure}[!thb]
\includegraphics[width=0.45\textwidth]{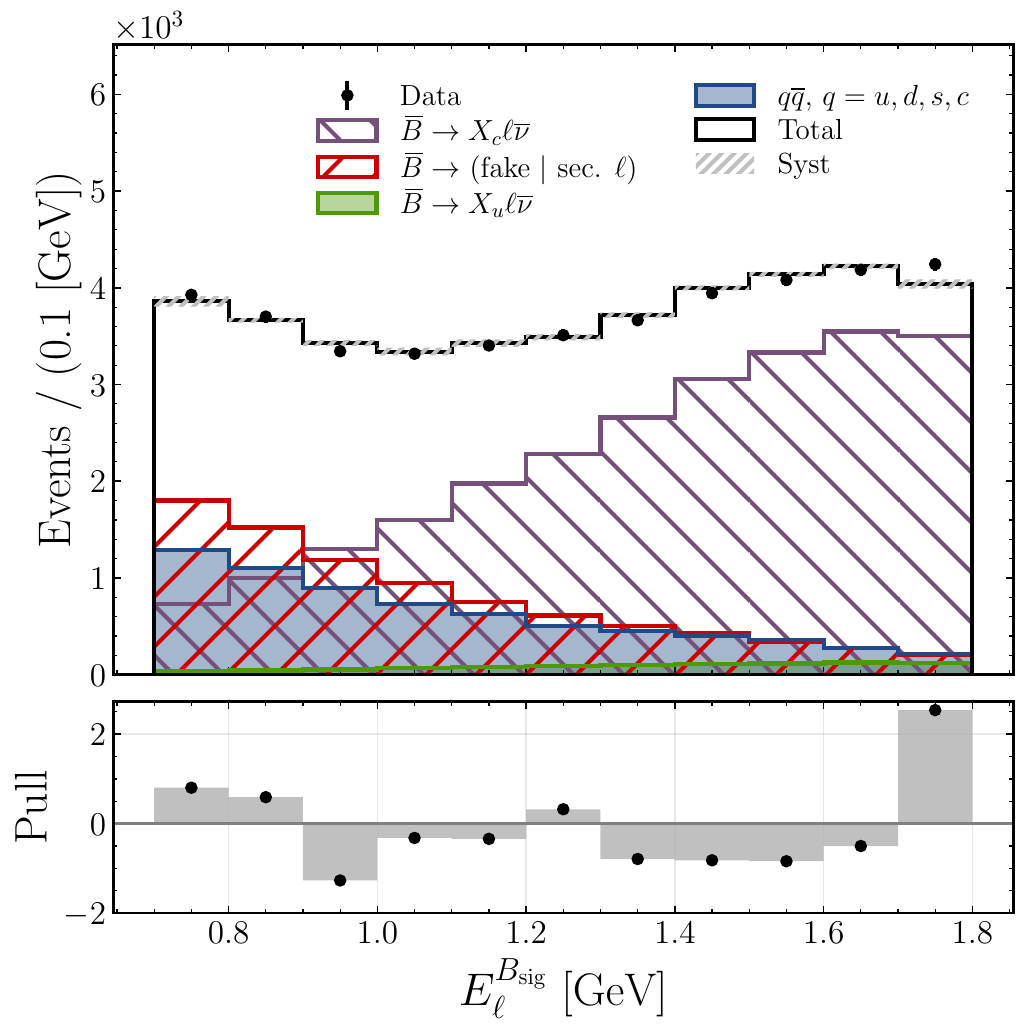} \\ 
\includegraphics[width=0.45\textwidth]{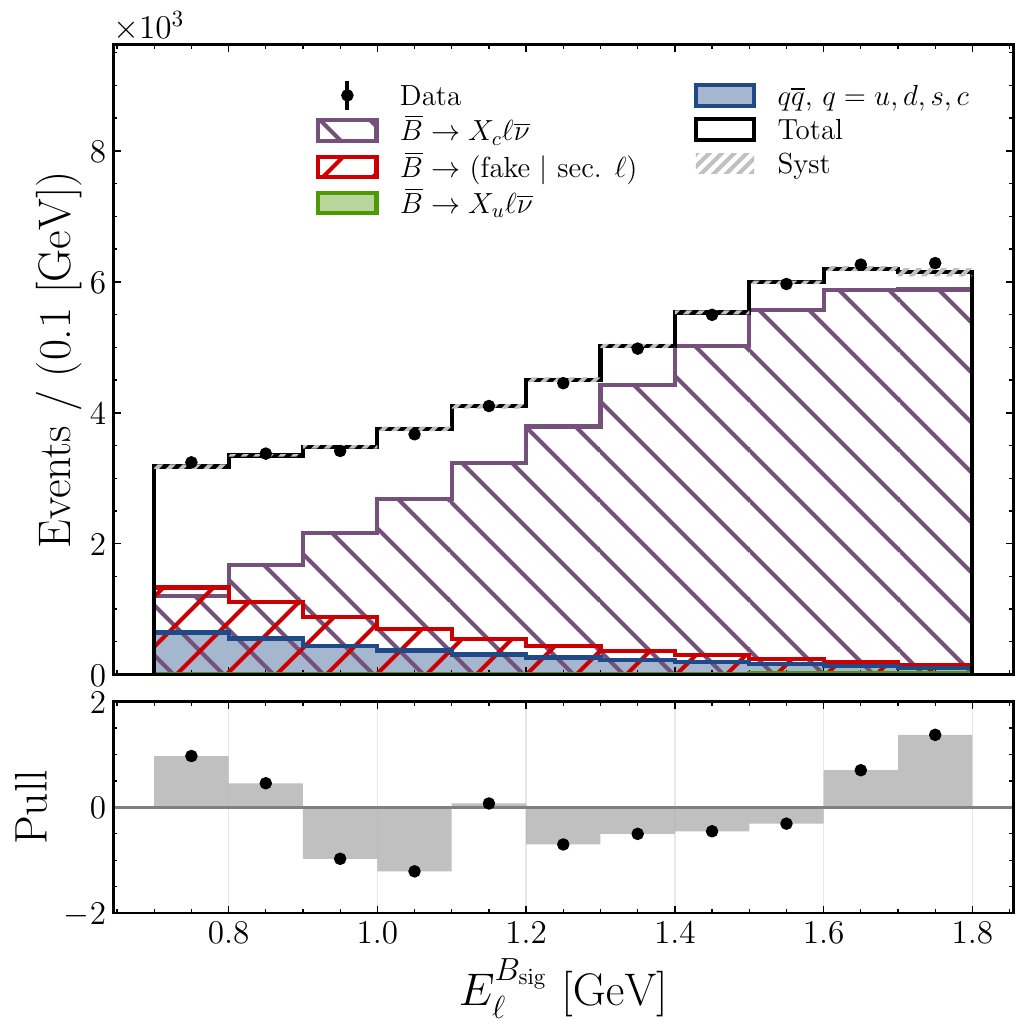} \\

\caption{\label{fig:sec_fakes_norm_fit}Fit to the secondary and fake lepton control region ($0.7 < \lepp < 1.8 \gev, M_X > 2.0 \gev$) in the $\BXulnu$ enhanced (\textit{top}) and depleted (\textit{bottom}) sub-samples for the $\BXulnu$ extraction sample. The four components are: correctly reconstructed $\BXulnu$ events (\textit{green}), correctly reconstructed $\BXclnu$ events (\textit{purple}), continuum events (\textit{blue}), and events in which either a hadron has been misidentified as a lepton or the lepton originates from a secondary decay (\textit{red}).}
\end{figure}

For each sub-sample we calibrate the contribution of non-continuum backgrounds by fitting a sample defined by the selection $0.7 < \lepp < 1.8 \gev, M_X > 2.0 \gev$, which partially overlaps with the signal region. This sample is enriched in backgrounds originating from: leptons from a secondary decay of the hadronic system, leptonic decays of tau-leptons from semitauonic events $\Bbar \to X \tau( \to \ell \nu \nub) \nub$, and purely hadronic decays where a hadron has been incorrectly identified as a lepton. These three backgrounds have similar shapes in the kinematic variables of interest and are thus grouped into a single component, secondary and fake leptons. For each sub-sample we perform a binned maximum likelihood fit in $\lepp$ floating the fraction of events assigned to the secondary and fake lepton or $\BXclnu$ components. The continuum component and $\BXulnu$ component, which is heavily suppressed in this region of phase-space, are kept fixed to expectation. The post-fit distributions are shown in Fig. \ref{fig:sec_fakes_norm_fit}. For the remainder of this work we scale the yield of the combined MC secondary and fake lepton contribution by the ratio of the fitted yield to the expectation in MC. The fits to the $\BXulnu$ enhanced (depleted) samples have a $\chi^2/ndf = 10.9/9 (4.5/9)$, respectively, where 

\begin{equation}
\chi^2 = \sum_{i,j} (N^{\rm obs}_i - N^{\rm exp}_i)(C^{\rm obs}_{\rm stat} + C^{\rm exp}_{\rm syst})^{-1}_{ij}(N^{\rm obs}_j - N^{\rm exp}_j).
\end{equation}
For details of the evaluation of $C^{\rm exp}_{\rm syst}$ see section \ref{sec:syst}. 

\begin{figure}[h]
    \centering
    \includegraphics[width=0.45\textwidth]{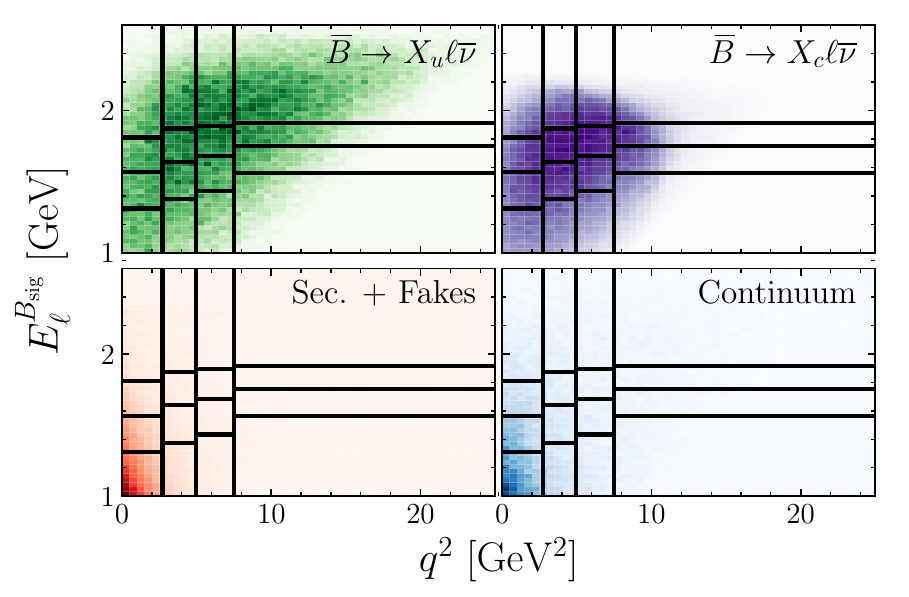}
    \caption{Binning structure for two dimensional fit to $q^2 : \lepp$. The sample is first divided into four equal $\BXclnu$ frequency $q^2$ bins. Each bin is then subdivided into four equal frequency $\lepp$ bins.}
    \label{fig:binning}
\end{figure}

We extract the $\BXulnu$ yield by means of a two-dimensional binned fit in $q^2: \lepp$ to the $\BXulnu$ enhanced sub-sample. The binning is chosen such that each bin is expected to have equal $\BXclnu$ yield. This ensures a large sample of $\BXclnu$ events are present in each bin allowing for use of a data-driven $\BXclnu$ template. A large fraction of the $\BXulnu$ events collect in the final broad $\lepp$ and $q^2$ bins, $q^2 > 7.5 \gevv, \lepp \gtrsim 1.8 \gev$ reducing the exposure to $\BXulnu$ modeling in the endpoint region. This binning is demonstrated in Fig. \ref{fig:binning} for the four components under consideration (secondary and fake leptons, continuum, $\BXulnu$, $\BXclnu$).

We derive a data-based template ($T$) for the $\BXclnu$ contribution from the $\BXulnu$ depleted sub-sample as 
\begin{equation}
    T_i = \tau_i (N^{\rm Data}_{i, D} - a \eta^{\Xulnu}_{i, D} - \eta^{\qqbar}_{i, D} - \eta^{\rm Sec.Fakes}_{i,D}), \label{eqn:T}
\end{equation}
where $N^{\rm Data}_{i,D}$ is the data yield in bin $i$ for the depleted sample and $\eta^{\Xulnu}_{i, D},  \eta^{\qqbar}_{i, D}, \eta^{\rm Sec.Fakes}_{i,D}$ are the MC yields for the $\BXulnu$, $\qqbar$, and secondary and fake lepton components, respectively. The coefficient $a$ is initially set to $1$; $\tau_i$ is a transfer factor from the $\BXulnu$ depleted to enhanced sub-sample given by the ratio of MC expectations in the enhanced and depleted samples, $\eta^{\Xclnu}_{i, E}$ and $\eta^{\Xclnu}_{i, D}$, respectively,
\begin{equation}
    \tau_i = \frac{\eta^{\Xclnu}_{i, E}}{\eta^{\Xclnu}_{i, D}},  
\end{equation}
and incorporates expected differences in the shape of the signal enhanced and depleted spectra, including different lepton energy and momentum transfer squared dependent efficiencies of the $|M^2_{\textrm{Miss}}|$ selection. 
The number of expected events in each bin of the enhanced sample is then given by

\begin{align}
H_i =&  N^{\rm Data}_{E} \Large[ \frac{\eta^{\rm Sec.Fakes}_{E}}{N^{\rm Data}_{E}}h^{\rm Sec.Fakes}_{i,E} \nonumber \\ 
&+ \frac{\eta^{\qqbar}_{E}}{N^{\rm Data}_{E}}h^{\textrm{\qqbar}}_{i,E} + f^{\Xulnu}_{E}h^{\Xulnu}_{i,E} \nonumber \\  &+ 
 (1 - f^{\Xulnu}_{E} - \frac{\eta^{\textrm{Sec.Fakes}}_{E} + \eta^{\qqbar}_{E}}{N^{\rm Data}_{E}}) \frac{T_i}{\sum_j T_j} \Large],
\end{align}

where $h_{i,E}^{\Xulnu}, h^{\textrm{\qqbar}}_{i,E}, h^{\rm Sec.Fakes}_{i,E}$ are the fraction of events of the $\BXulnu, \qqbar$, and secondary and fake lepton components reconstructed in bin $i$, respectively, as determined by the MC simulation. The total data yield is given by $N^{\rm Data}_{E}$, and $\eta^{\qqbar}_{E}, \eta^{\rm Sec.Fakes}_{E}$ are the expected yield of the continuum and secondary and fake lepton component in the $\BXulnu$ enhanced sample, respectively. The parameter $f^{\Xulnu}_{E}$ is the fraction of events assigned to the $\BXulnu$ component and is floated in the fit. 

\begin{figure*}[htb]
    \centering
    \includegraphics[width=0.8\textwidth]{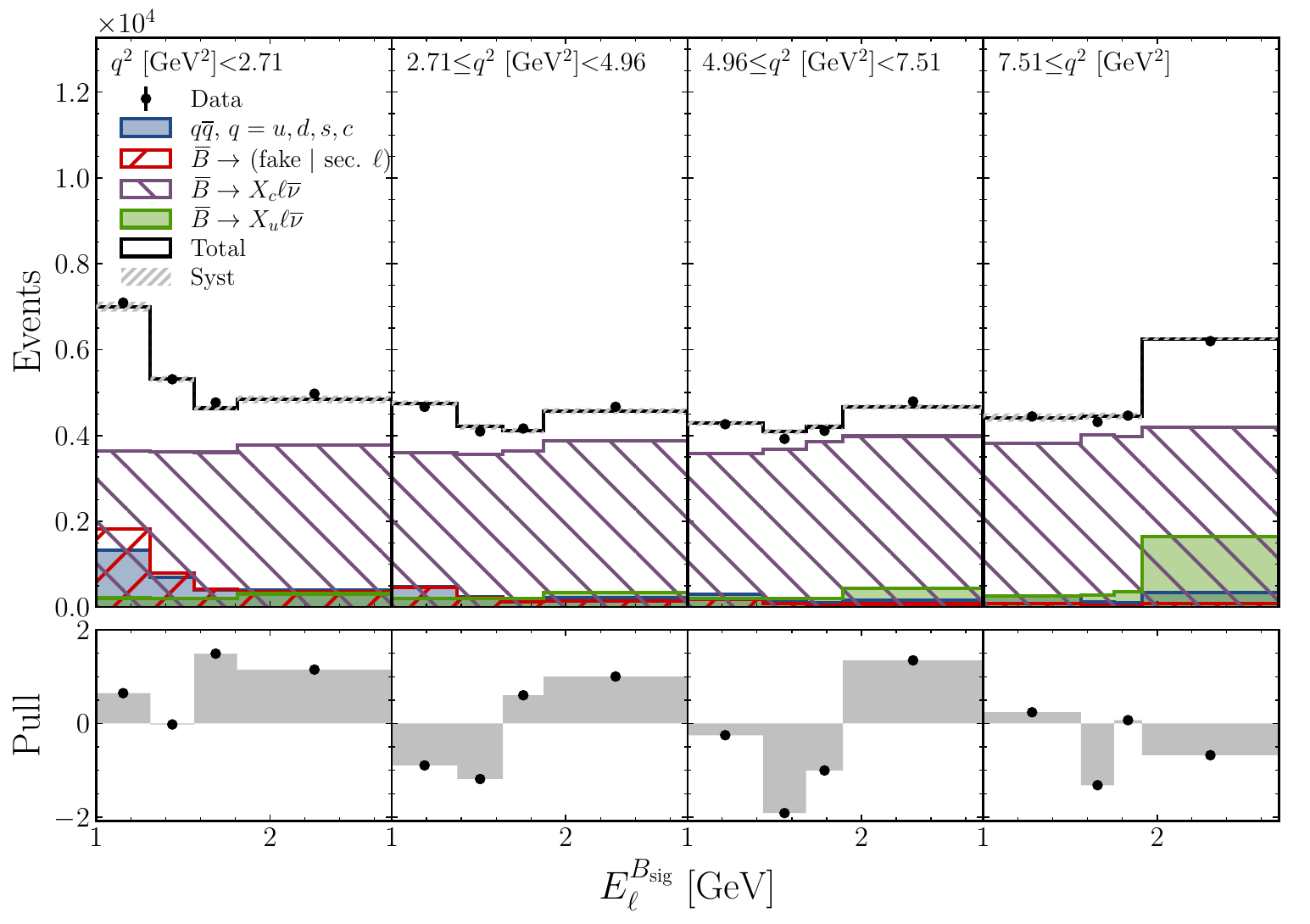}
    \caption{Fit to the $q^2 : \lepp$ distribution. The four components are: correctly reconstructed $\BXulnu$ events (\textit{green}), correctly reconstructed $\BXclnu$ events (\textit{purple}), continuum events (\textit{blue}), and events in which either a hadron has been misidentified as a lepton or the lepton originates from a secondary decay (\textit{red}).}
    \label{fig:xulnu_fit}
\end{figure*}

From the MC simulation we expect $7\%$ of $\BXulnu$ events to contaminate the $\BXulnu$ depleted sample. To reduce the dependence on the assumed $\BXulnu$ branching fraction in MC we repeat the fit $20$ times updating the coefficient $a$ as
\begin{equation}
    a = \frac{f^{\Xulnu}_{E} N^{\rm Data}_{E}}{\eta^{\Xulnu}_{E}}, 
\end{equation} on each iteration. Convergence is observed within $4$ iterations. The result of the fit is shown in Fig. \ref{fig:xulnu_fit}. The fit yields $N^{\Xulnu} = f^{\Xulnu}_{E} N^{\rm Data}_{E} =  5430 \pm 450_{\rm stat} \pm 350_{\rm syst}$ $\BXulnu$ events with a $\chi^2/{\rm ndf} = 15.1/14$. 

\subsection{\texorpdfstring{$\BXclnu$}{Charmed Semileptonic B Decay} Yield}
\begin{figure}[!thb]
\includegraphics[width=0.45\textwidth]{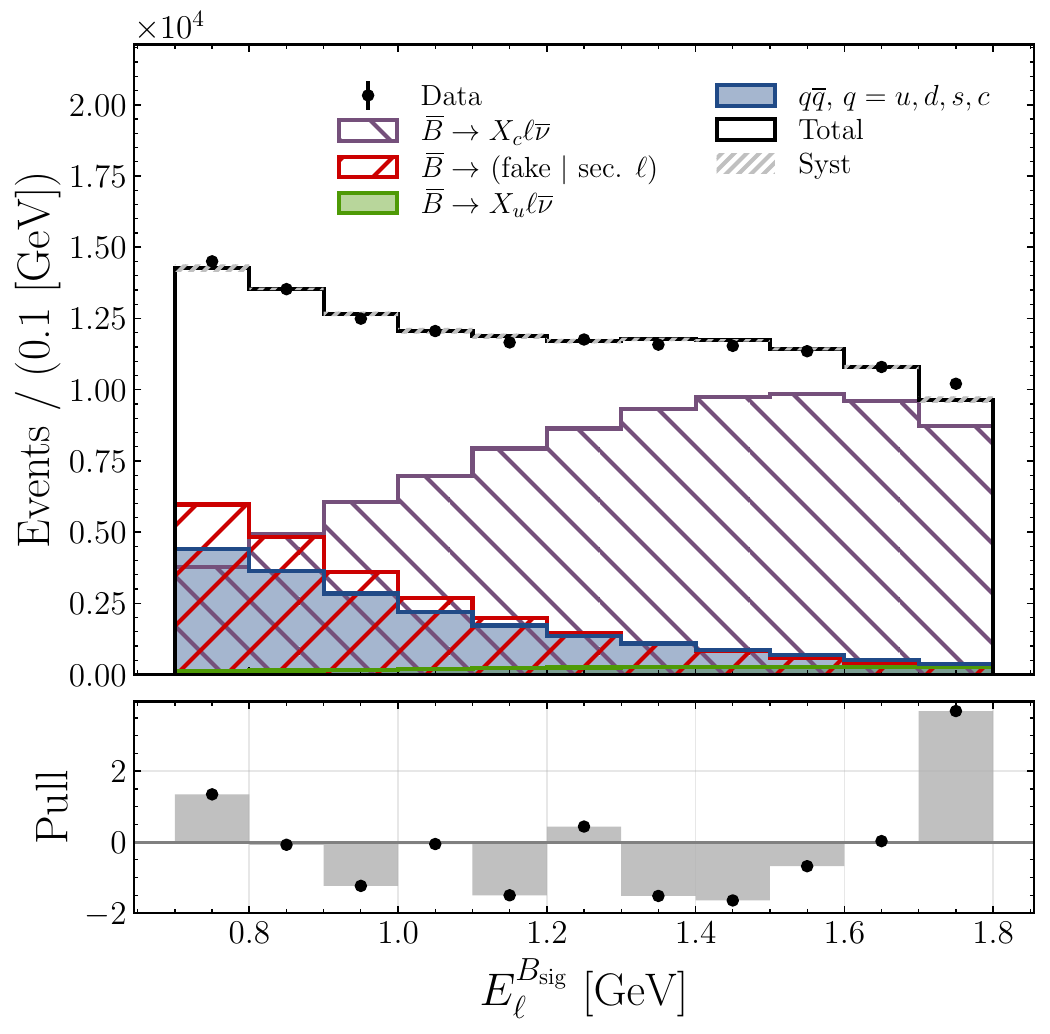} \\ 
\includegraphics[width=0.45\textwidth]{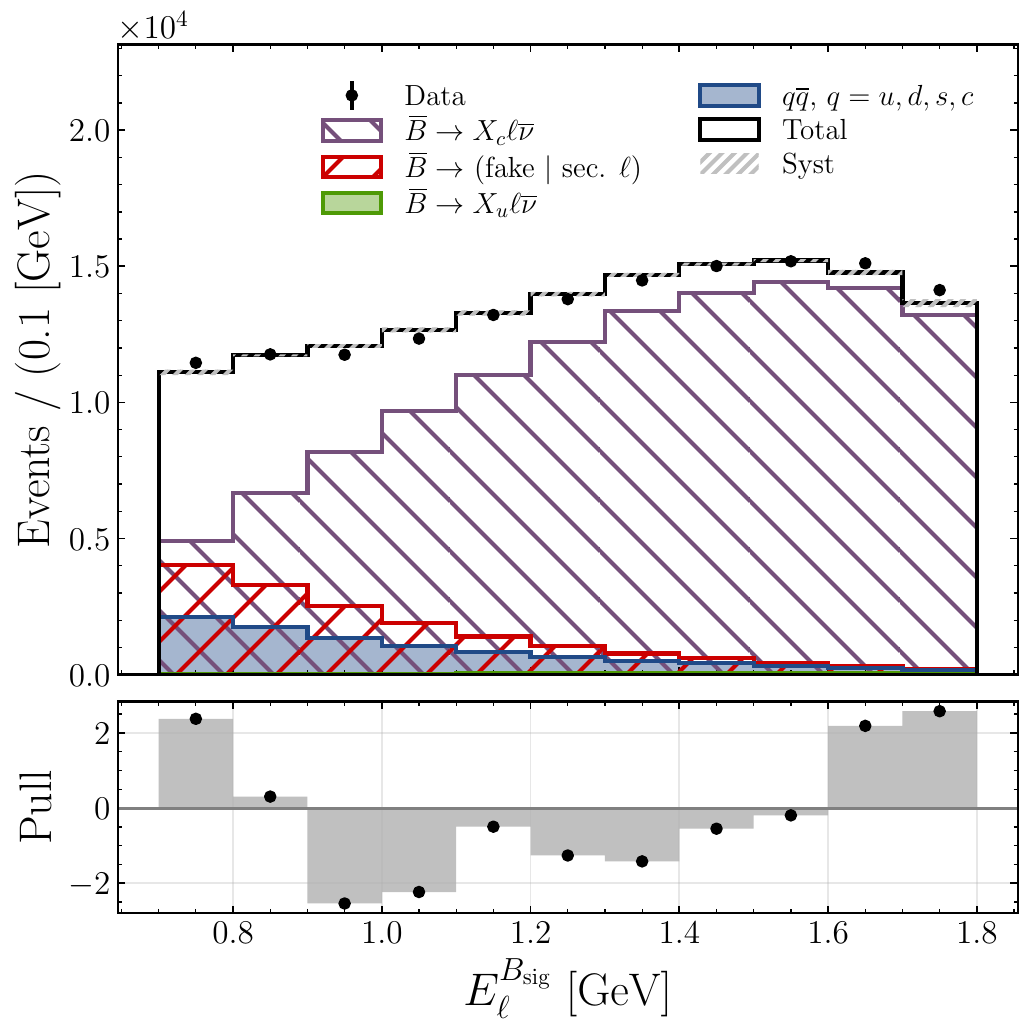} \\
\caption{Fit to the secondary and fake lepton control region ($0.7 < \lepp < 1.8 \gev, M_X > 2.0 \gev$) in the $\BXulnu$ enhanced (\textit{top}) and depleted (\textit{bottom}) sub-samples for the $\BXclnu$ extraction sample.}
\label{fig:sec_fakes_norm_fit_xclnu}
\end{figure}

\begin{figure}[h]
    \centering
    \includegraphics[width=0.45\textwidth]{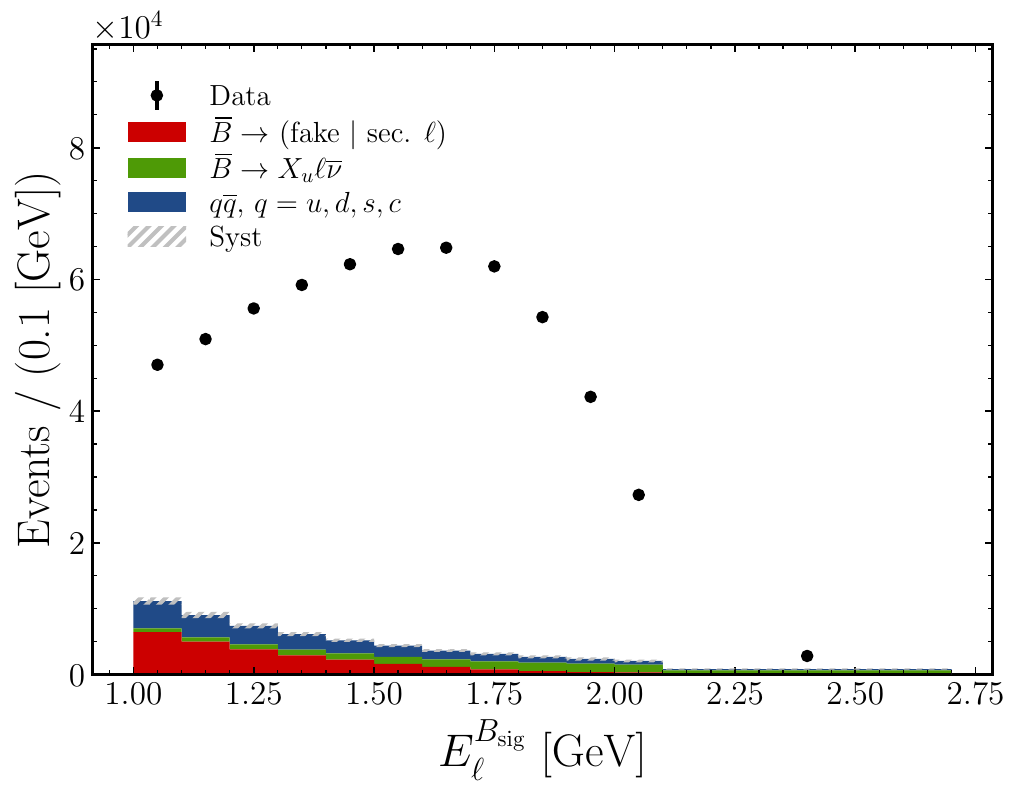}
    \caption{The determined background compared with all events in the $\BXclnu$ extraction sample.}
    \label{fig:sampleA_lepp}
\end{figure}

To extract the $\BXclnu$ yield we broaden our selection, removing the $D^*$ veto and $|M^2_{\textrm{Miss}}|$ requirements.
The secondary and fake lepton component is normalized following the procedure established for the $\BXulnu$ extraction sample. The fits are presented in Fig.~\ref{fig:sec_fakes_norm_fit_xclnu} and have $\chi^2/{\rm ndf} = 24.1/9 (16.0/9)$ for the enhanced (depleted) sub-samples. These large $\chi^2/{\rm ndf}$ are dominated by the contribution from the final bin in $\lepp$. To test the impact of this mismodeling we repeat the whole procedure removing the final bin from all four secondary and fake lepton normalization fits, finding $\chi^2/{\rm ndf} = 2.8/8 (3.4/8)$ and $7.1/8 (14.0/8)$ for the enhanced (depleted) sub-samples of the $\BXulnu$ and $\BXclnu$ extraction samples, respectively. The central value of $\Delta \mathcal{B}(\BXulnu)/ \Delta \mathcal{B}(\BXclnu)$ shifts by $+0.5\%$. The reconstructed $\lepp$ spectrum of the combined enhanced and depleted sub-samples is shown in Fig. \ref{fig:sampleA_lepp}. Given the high purity of the sample ($>90\%$), the $\BXclnu$ yield is found via a simple background subtraction: 
\begin{equation}
    N^{\Xclnu} =  N^{\rm Data} - \eta^{\qqbar} - \eta^{\rm Sec.Fakes} - a\eta^{\Xulnu},
\end{equation}
where $N^{\rm Data}$ is the data yield in the broadened selection, and $\eta^{\qqbar} ,\eta^{\rm Sec.Fakes},$ and $\eta^{\Xulnu}$ are the MC expectations of the continuum, secondary and fake lepton background, and $\BXulnu$ component.

The ratio of partial branching fractions is given by 
\begin{equation}
    \frac{\Delta \mathcal{B}(\BXulnu)}{\Delta \mathcal{B}(\BXclnu)} = \frac{\epsilon^{\Xclnu} N^{\Xulnu}}{\epsilon^{\Xulnu} N^{\Xclnu}},    
\end{equation}
where $\epsilon^{\Xulnu}$, $\epsilon^{\Xclnu}$ are the reconstruction efficiencies of the $\BXulnu$ and $\BXclnu$ events, respectively, and are estimated from MC simulations.  The results are presented in Section \ref{sec:results}.

To validate the fit procedure we generate ensembles of pseudo-experiments for different input branching fractions for $\BXulnu$ and $\BXclnu$ decays. No bias in central values is observed. 

\section{\label{sec:syst} Systematic Uncertainties}

Several systematic uncertainties affect the measured ratio of partial branching fractions. The most important sources of systematic uncertainties are associated with the modeling of the $\BXulnu$ component and the composition of the secondary and fake lepton component. Each systematic effect is varied independently and the analysis procedure repeated. All systematic uncertainties are taken as uncorrelated and summed in quadrature for the total systematic uncertainty.

\subsection{\texorpdfstring{$\BXulnu$}{Charmless Semileptonic B Decay} Modeling}

As the simulation of $\BXulnu$ events is a hybrid composition of low-mass resonant and high-mass non-resonant states the relative contributions of the different states will impact the reconstruction efficiency, and shape of the $\BXulnu$ template. We evaluate the uncertainty by varying the assumed branching fractions of the resonant decays, $\Bbar \to (\pi, \rho, \omega, \eta, \eta^\prime) \ell \nub$ by one standard deviation, for $\Bbar \to (\pi, \rho, \omega) \ell \nub$ decays we vary the form-factors along the eigen-directions of the covariance matrix of their respective parameters. We assign the quadratic mean of the differences between the two varied results and the nominal result as the uncertainty.

For $\Bbar \to (\eta, \eta^\prime) \ell \nub$ decays we replace the nominal model with the alternate description of Ref.~\cite{Ball:2007hb} and assign the full difference as the uncertainty. For the non-resonant channels we vary $m_b^{\rm KN}$ and $a^{\rm KN}$ by their uncertainties in the eigen-directions of their covariance matrix, we investigate the impact of reweighting the DFN model to the model of Bosch, Lange, Neubert, and Paz (BLNP) \cite{Lange:2005yw} with $b$ quark mass in the shape-function scheme $m_b^{\textrm{SF}} = 4.61 \gev$ and $\mu_\pi^{2\, \textrm{SF}} = 0.2 \gevv$ and consider the full difference to the nominal result as the uncertainty. We vary the assumed $\BXulnu$ yield by $\pm 1 \sigma$. For each variation investigated the hybrid weights are recalculated following Eq. \ref{eq:hybrid}. 

The cross-feed fraction of $\BXulnu$ events into the $\BXulnu$ depleted sample depends on the production rate of $K^+$ and $\KS$ in the fragmentation of the $X_u$ system. We vary the relative weight of simulated $\BXulnu$ events generated with kaon pairs by $\pm 25\%$.

\subsection{\texorpdfstring{$\BXclnu$}{Charmed Semileptonic B Decay} Modeling}
We vary the values of the branching fractions of $\Bbar \to D^{(*(*))} \ell \nub$ about the nominal value by MC simulation, assuming a Gaussian error profile. For the unmeasured gap channels, $\Bbar \to D^{**}_{\textrm{Gap}} ( \to D^{(*)} \eta ) \ell \nub$, we assume a uniform error profile between zero and twice the nominal branching fraction. For both cases the uncertainty is taken to be half the difference in the values that correspond to the 15.9 and 84.1 percentiles of the obtained distribution. We vary the form factors up and down by one standard deviation along the eigen-directions of the covariance matrix. 

As we do not reconstruct $\KL$ mesons, the momentum they carry away is attributed to the neutrino, biasing the reconstructed $q^2$ towards higher values. Charmed semileptonic events in which a $\KL$ is produced within the hadronic system are predominantly assigned to the $\BXulnu$ enhanced sample, causing a small bias in the data-driven template correction.
We estimate the impact of this bias via an in-situ calibration of the $\BXclnu$ $q^2$ spectra. The $\BXclnu$ background in the $\BXulnu$ depleted sub-sample is unfolded in $q^2$ after background subtraction, following the procedure established in Sec~\ref{sec:unfolding}. A fourth-order Chebyshev polynomial is fit to the ratio of the normalized unfolded data yield to the normalized MC yield. Correction weights are applied to the $\BXclnu$ simulated events during the $\BXulnu$ extraction by evaluating the polynomial at the generated $q^2$ of each event. The full difference to the nominal result is taken as a systematic uncertainty.

\subsection{Other}

Uncertainties associated with particle identification: lepton identification efficiencies, hadron to lepton misidentification rates, kaon identification efficiency, pion to kaon fake rates, $\KS$, and slow pion efficiency, are varied within their uncertainty. The impact of tracking performance is estimated by varying the weight of events by $0.35\%$ per track on the signal side.  The uncertainty on the off-resonance calibration of the continuum contribution is estimated by varying the coefficients of the linear calibration functions within their uncertainty.

The uncertainty on the shape of the secondary and fake lepton contributions is estimated by varying the normalization of the three components in simulation: semi-tauonic decays, events with leptons from a secondary decay, and events where a hadron has been misidentified. A conservative $30\%$ normalization uncertainty is assigned to each component.

The number of $\BB$ pairs in the sample \cite{Belle:2012iwr}, and the fraction of $\Y4S$ mesons that decay to charged $\BpBm$ pairs are varied within their uncertainty \cite{ParticleDataGroup:2020ssz}.
The uncertainty due to the limited size of the MC samples used is estimated via a bootstrapping method. 

Systematic variations that affect the signal side normalization and the data statistical uncertainty are fully propagated to the tag calibration. Uncertainty due to the tag calibration is therefore not separately identified but rather included in the already discussed sources.

\subsection{Statistical Uncertainty and Correlations}

The statistical uncertainty and correlations are determined using a Poisson bootstrapping procedure \cite{chamandyEstimatingUncertaintyMassive2012}. Ensembles of the collision data-set are created and the entire analysis procedure is repeated, beginning with the calibration of the tagging algorithm. The correlation between observables is estimated as the Pearson correlation coefficient from the central values of the trials.

\subsection{Systematic Correlations}

For each measurement under consideration, the correlation of systematic uncertainties between bins or between measurements is estimated by analyzing the individual pseudo-experiment studies. For systematic error sources estimated as the difference between a nominal and alternate model, we consider all bins to be fully correlated or fully anti-correlated. The covariance for these sources is then given by
\begin{equation}
    C = \Sigma J \Sigma,
\end{equation}
where $J$ is a matrix-of-ones, $\Sigma = \textrm{diag}(\vec{\sigma})$, and $\vec{\sigma}$ is the vector of estimated uncertainties in the bins or measurements under consideration. For systematic sources whose impact is evaluated by varying an input by one standard deviation about the nominal value, we average the covariance matrix estimated in this manner for both variations.

For systematic uncertainties estimated via the distribution of pseudo-experiment samples, we extract the correlation from the samples. The total systematic covariance $C_{\rm syst}^{\rm exp}$ is given by
\begin{equation}
    C_{\rm syst}^{\rm exp} = \sum_x C^x_{\rm syst},
\end{equation}
where $x$ runs over the individual sources of systematic uncertainty under consideration.

\section{\label{sec:results} Partial Branching Fraction Results}
The measured ratio of partial branching fractions ($R$) with detailed breakdown of the systematic uncertainties is given in Table \ref{tab:ratio_results}.

\begin{table}
\caption{\label{tab:ratio_results} Summary of the central value ($R$), statistical, and systematic uncertainties for the ratio of partial branching fractions. The uncertainties are given as relative values on the central value in percent. 
}
\begin{ruledtabular}
\begin{tabular}{@{\hspace{3em}} l r @{\hspace{3em}}}
$R \times 100 $ & $1.99$\\ \hline
Stat. Error (Data) & $8.5$\\ \hline
$\mathcal{{B}}(\Bbar \to \pi/\eta/\rho/\omega/\eta^\prime \ell \nub)$ & $0.2$\\ 
$\mathcal{{FF}}(\Bbar \to \pi/\eta/\rho/\omega/\eta^\prime \ell \nub)$ & $0.3$\\ 
$\mathcal{{B}}(\Bbar \rightarrow x_{u} \ell \nub)$ & $0.8$\\ 
Hybrid Model (BLNP) & $1.0$\\ 
DFN ($m_b^{\rm KN}, a^{\rm KN}$) & $5.1$\\ 
$N_{g \to s\overline{s}}$ & $0.4$\\   \hline
$\mathcal{{B}}(\Bbar \to D \ell \nub)$ & $0.1$\\ 
$\mathcal{{B}}(\Bbar \to D^{*} \ell \nub)$ & $0.8$\\ 
$\mathcal{{B}}(\Bbar \to D^{**} \ell \nub)$ & $0.3$\\ 
$\mathcal{{B}}(\Bbar \to D^{{(*)}} \eta \ell \nub)$ & $0.2$\\ 
$\mathcal{{B}}(\Bbar \to D^{{(*)}} \pi \pi \ell \nub)$ & $0.2$\\ 
$\mathcal{FF}(\Bbar \to D \ell \nub)$ & $0.2$\\ 
$\mathcal{FF}(\Bbar \to D^{*} \ell \nub)$ & $0.9$\\ 
$\mathcal{{FF}}(\Bbar \to D^{**} \ell \nub)$ & $0.4$\\ 
Sec.Fakes. Composition & $3.8$\\ 
In-situ $q^2$ Calibration & $2.8$\\  \hline
$\ell$ID Efficiency & $0.1$\\ 
$\ell$ID Fake Rate & $0.3$\\ 
$K\pi$ID Efficiency & $1.0$\\ 
$K\pi$ID Fake Rate & $0.6$\\ 
$K_{S}^{0}$ Efficiency & $0.2$\\ 
$\pi_{\rm slow}$ Efficiency & $<0.1$\\ 
Tracking & $0.1$\\ 
Continuum Calibration & $0.4$\\ 
$N_{{BB}}$ & $<0.1$\\ 
$f_{+/0}$ & $<0.1$\\ 
Stat. Error (MC)  & $2.6$\\ \hline
Total Syst. & $7.8$\\ 
\end{tabular}
\end{ruledtabular}
\end{table}

We test the expectation of lepton flavor universality by repeating the analysis procedure separately for the electron and muon modes. Systematic uncertainties are evaluated by taking into account correlations between the lepton modes. The goodness of fit is $\chi^2/{\rm ndf} = 15.3/14 (12.7/14)$ for the electron (muon) mode $\BXulnu$ extraction fit, respectively. We obtain

\begin{align}
    \frac{\Delta \mathcal{B}(\Bbar \to X_u e\nub)}{\Delta \mathcal{B}(\Bbar \to X_c e \nub)} &= (1.74 \pm 0.24_{\rm stat} \pm 0.14_{\rm syst}) \times 10^{-2}, \\
    \frac{\Delta \mathcal{B}(\Bbar \to X_u \mu \nub)}{\Delta \mathcal{B}(\Bbar \to X_c \mu \nub)} &= (2.21 \pm 0.24_{\rm stat} \pm 0.21_{\rm syst}) \times 10^{-2},
\end{align}

\noindent where the expected cross-feed is negligible. We further compare directly the efficiency corrected yields of the charmed and charmless mode and find

\begin{align}
    \frac{\Delta \mathcal{B}(\Bbar \to X_u e \nub)}{\Delta \mathcal{B}(\Bbar \to X_u \mu \nub)} &= 0.79 \pm 0.13_{\rm stat} \pm 0.07_{\rm syst}, \\
    \frac{\Delta \mathcal{B}(\Bbar \to X_c e \nub)}{\Delta \mathcal{B}(\Bbar \to X_c \mu \nub)} &= 1.002 \pm 0.005_{\rm stat} \pm 0.024_{\rm syst}, 
\end{align}

where the systematic uncertainty in the $\BXclnu$ mode is dominated by lepton identification efficiency uncertainties and in the $\BXulnu$ mode by the composition of the secondary and fake lepton component and MC statistics. The normalization of hadrons misidentified as leptons has been varied independently for the electron and muon channels. The charmless mode is compatible with unity within $1.4$ standard deviations. We extrapolate the charmed mode to the full phase-space via a correction factor of $1.0045\pm 0.0001$, extracted from simulation, finding 
\begin{equation}
    \frac{\mathcal{B}(\Bbar \to X_c e \nub)}{\mathcal{B}(\Bbar \to X_c \mu \nub)} = 1.007 \pm 0.005_{\rm stat} \pm 0.025_{\rm syst},
\end{equation}

which is in excellent agreement with the SM prediction, $1.006\pm0.001$ \cite{Rahimi:2022vlv}, and a recent measurement with \belletwo\ data that found $1.007\pm 0.009_{\rm stat} \pm 0.019_{\rm syst}$ \cite{Belle-II:2023qyd}.

Isospin breaking effects, such as weak annihilation, can be constrained by measuring the ratio of partial branching fractions separately for charged and neutral $\B$ mesons \cite{Belle:2021idw}. We measure the double ratio
\begin{equation}
    R_{\textrm{iso}} = \frac{\Delta \mathcal{B}(B^+ \to X_u^0 \ell^+ \nu)}{\Delta \mathcal{B}(B^+ \to X_c^0 \ell^+ \nu)} \times \frac{\Delta \mathcal{B}(B^0 \to X_c^- \ell^+ \nu)}{\Delta \mathcal{B}(B^0 \to X_u^- \ell^+ \nu)}. 
\end{equation}
Isospin breaking effects in the charmed mode are expected to be negligible, with $\Gamma(B^0 \to X_c^- \ell^+ \nub) = \Gamma(B^+ \to X_c^0 \ell^+ \nu)$ up to $\mathcal{O}(1/m_b^3)$ \cite{Gronau:2006ei}. Taking this result, the above relation reduces to
\begin{equation}
    R_{\textrm{iso}} = \frac{\tau_{B^0}}{\tau_{B^+}}\frac{\Delta\mathcal{B}(B^+ \to X_u^0 \ell^+ \nu)}{\Delta\mathcal{B}(B^0 \to X_u^- \ell^+ \nu)}. 
\end{equation}

The measured number of $\BXulnu$ and $\BXclnu$ events in the neutral and charged channels are corrected for cross-feed due to charge misassignment by the $B$-tagging algorithm, with the probability of such misassignment taken from simulation. We find

\begin{align}
    \frac{\Delta \mathcal{B}(B^+ \to X_u \ell^+ \nu)}{\Delta \mathcal{B}(B^+ \to X_c \ell^+ \nu)} &= (1.76 \pm 0.22_{\rm stat} \pm 0.17_{\rm syst})\times 10^{-2}, \\
    \frac{\Delta \mathcal{B}(B^0 \to X_u \ell^+ \nu)}{\Delta \mathcal{B}(B^0 \to X_c \ell^+ \nu)} &= (2.53 \pm 0.34_{\rm stat} \pm 0.22_{\rm syst})\times 10^{-2}, \\
    R_{\rm iso} =& 0.70 \pm 0.13_{\rm stat} \pm 0.07_{\rm syst},
\end{align}

\noindent where the dominant systematic on $R_{\rm iso}$ is due to the MC statistical error and we find $\chi^2/{\rm ndf} = 13.8/14$, and $16.2/14$ for the charged and neutral $B$ sample $\BXulnu$ extraction fits, respectively. This value of $R_{\rm iso}$ is $2.0\sigma$ apart from the expectation of equal semileptonic rates for both isospin states. The relative contribution of weak annihilation to the total can be defined as 
\begin{equation}
    \frac{\Gamma_{WA}}{\Gamma} = \frac{f_u}{f_{WA}}(R_{\rm iso}-1), 
\end{equation}
where $\Gamma$ is the total $\BXulnu$ decay width and $f_u$, $f_{WA}$ give the fraction of phase space considered in the measurement of $R_{iso}$ for $\BXulnu$ and weak annihilation, respectively. Weak annihilation is expected to be confined to the high $q^2$ region, $q^2 \approx m_B^2$, we therefore assume $f_{WA} \approx 1$. The value of $f_u$ is extracted from MC to fully propagate shape varying uncertainties. The central value is $f_u = 0.87$ following the prescription of the DFN model \cite{DeFazio:1999ptt} utilized in the construction of the hybrid MC. We find $-0.54 < \frac{\Gamma_{WA}}{\Gamma}  < -0.02$ at $90\%$ confidence level. This is consistent with recent results from Belle and BaBar which set 90\% confidence levels at $-0.14 < \frac{\Gamma_{WA}}{\Gamma}  < 0.17$ \cite{Belle:2021eni} and $-0.17 < \frac{\Gamma_{WA}}{\Gamma} < 0.19$ \cite{BaBar:2011xxm}, respectively. The result can also be compared with the strongest experimental constraints on weak annihilation that come from a model-dependent direct measurement of the $q^2$ spectrum, $\frac{\Gamma_{WA}}{\Gamma} < 7.4\%$ \cite{CLEO:2006xbj} as well as theoretical estimates of the relative rate extrapolated from a study of $D$ and $D_s$ decays, $\frac{\Gamma_{WA}}{\Gamma} \sim 1-3\%$ \cite{Bigi:1993bh, Ligeti:2010vd, Gambino:2010jz, Voloshin:2001xi}.

\section{\texorpdfstring{$|V_{ub}|/|V_{cb}|$}{|Vub|/|Vcb|} and \texorpdfstring{$|V_{ub}|$}{|Vub|} determination}
\label{sec:vub}
There are no calculations of the ratio of partial rates that take into account the correlations in the input parameters, which would likely lead to partial cancellation of the uncertainties. We therefore perform a naive extraction of $|V_{ub}|/|V_{cb}|$ and $|V_{ub}|$.
The relationship between the measured ratio of partial branching fractions and the ratio of CKM element magnitudes is given by
\begin{equation}
    \frac{|\Vub|}{|\Vcb|} = \sqrt{\frac{\Delta \mathcal{B}(\BXulnu)}{\Delta \mathcal{B}(\BXclnu)}\frac{\Delta\Gamma(\BXclnu)}{\Delta\Gamma(\BXulnu)}},
\end{equation}
where $\Delta\Gamma(\BXulnu)$ and $\Delta\Gamma(\BXclnu)$ are the predictions of the partial rates for lepton energies above $1 \gev$, omitting the CKM factors, of $\BXulnu$ and $\BXclnu$ respectively. \\

For $\BXulnu$, we consider two predictions of the partial rate. Those of BLNP \cite{Lange:2005yw}, described in section \ref{sec:syst}, $\Delta\Gamma^{\rm BLNP}(\BXulnu) = 61.5^{+6.4}_{-5.1} \ps^{-1}$, and of Gambino, Giordano, Ossola, and Uraltsav (GGOU) \cite{Gambino:2007rp}, using as input $m_b^{\rm kin} = 4.55 \pm 0.02 \gev$ and $\mu_{\pi}^{2\,{\rm kin}} = 0.45 \pm 0.08 \gevv$, $\Delta\Gamma^{\rm GGOU}(\BXulnu) = 58.5^{+2.7}_{-2.3} \ps^{-1}$.

For $\BXclnu$, we extract the partial rate from the global fit to the moments of the $\BXclnu$ spectra in the kinetic scheme of Ref.~\cite{HFLAV:2019otj}, and estimate the phase-space correction factor from MC, $f_c = 0.784\pm0.005$, finding $\Delta\Gamma(\BXclnu)=29.7 \pm 1.2 \ps^{-1}$. We measure

\begin{align}
    \frac{|\Vub|}{|\Vcb|}^{\rm BLNP} &= (9.81 \pm 0.42_{\rm stat} \pm 0.38_{\rm syst}  \nonumber \\
        &{\hphantom{}} \pm 0.51_{ \Delta\Gamma(\BXulnu)} \pm 0.20_{ \Delta\Gamma(\BXclnu)}) \times 10^{-2}, \\
    \frac{|\Vub|}{|\Vcb|}^{\rm GGOU} &= (10.06 \pm 0.43_{\rm stat} \pm 0.39_{\rm syst} \nonumber \\
        &{\hphantom{}} \pm 0.23_{ \Delta\Gamma(\BXulnu)} \pm 0.20_{ \Delta\Gamma(\BXclnu)}) \times 10^{-2},
\end{align}

\noindent where the third and fourth uncertainties correspond to the uncertainty of the partial rate predictions for $\BXulnu$ and $\BXclnu$, respectively.
These values are in excellent agreement with the world averages of both the inclusive and exclusive determinations of $|\Vub|/|\Vcb|$ presented in section \ref{sec:introduction}. 
We also extract $|V_{ub}|$, with the relation
\begin{equation}
    |\Vub| = \sqrt{\frac{\Delta \mathcal{B}(\BXulnu)}{\Delta \mathcal{B}(\BXclnu)}\frac{\Delta \mathcal{B}^{\rm ex.}(\BXclnu)}{\tau_B \Delta\Gamma(\BXulnu)}},
\end{equation}
where $\tau_B = 1.579 \pm 0.004 \ps$ \cite{HFLAV:2019otj} is the average $B$ meson lifetime, and $\Delta \mathcal{B}^{ex.}(\BXclnu)$ is the externally measured partial branching fraction of $\BXclnu$. For the latter, we take the average of the values reported in Refs.~\cite{Belle:2006kgy, BaBar:2009zpz}, assuming them to be fully uncorrelated, finding
\begin{equation}
\Delta \mathcal{B}^{ex.}(\BXclnu) =  (8.55 \pm 0.13) \%.  
\end{equation}
We measure 
\begin{align}
    |\Vub|^{\rm BLNP} &= (4.19 \pm 0.18_{\rm stat} \pm 0.17_{\rm syst} \nonumber \\
        &{\hphantom{}} \pm 0.22_{\Delta\Gamma(\BXulnu)} \pm 0.03_{\Delta \mathcal{B}^{ex.}(\BXclnu)}) \times 10^{-3}, \\
    |\Vub|^{\rm GGOU} &= (4.29 \pm 0.18_{\rm stat} \pm 0.17_{\rm syst} \nonumber \\
        &{\hphantom{}} \pm 0.10_{\Delta\Gamma(\BXulnu)} \pm 0.03_{\Delta \mathcal{B}^{ex.}(\BXclnu)}) \times 10^{-3},
\end{align}

which are in excellent agreement with the inclusive world average value.

\section{\label{sec:unfolding} Unfolding Procedure}

In addition to the ratio of partial branching fractions we report the differential ratio of partial branching fractions in $q^2$ and $\lepp$ to allow for future model-independent determinations of $|V_{ub}|/|V_{cb}|$. The ratio of differential branching fractions is extracted by unfolding individually the $\BXulnu$ and $\BXclnu$ yields. The samples are projected into $\lepp$ and $q^2$ bins defined by the boundaries:

\begin{align*}
    \BXulnu : \\
    \lepp &: [1.0, 1.1, 1.2, 1.3, 1.4, 1.5, 1.6, 1.7,\\ &\hphantom{:]]]} 1.8, 1.9, 2.0, 2.1, 2.2, 2.3, 2.7] \gev, \\
    q^2 &: [0,2,4,6,8,10,12,26] \gevv, \\
    \BXclnu : \\
    \lepp &: [1.0, 1.1, 1.2, 1.3, 1.4, 1.5, 1.6, 1.7,\\ &\hphantom{:]]]} 1.8, 1.9, 2.0, 2.1, 2.7] \gev, \\
    q^2 &: [0,2,4,6,8,10,26] \gevv.
\end{align*}

All components are normalized to their respective yields during the $\BXulnu$ and $\BXclnu$ extraction described in section \ref{sec:measurement}. No additional selection criteria are applied. All backgrounds, including continuum,  secondary and fake leptons, and $\BXclnu$, or $\BXulnu$, respectively, for the $\BXulnu$ and $\BXclnu$ unfolding are then subtracted. For the $\BXulnu$ unfolding the shape of the $\BXclnu$ component is derived following Eq. \ref{eqn:T}. The statistical and systematic uncertainties and correlations are estimated following the description of Sec. \ref{sec:syst}. 

\begin{figure}[h]
    \centering
    \includegraphics[width=0.45\textwidth]{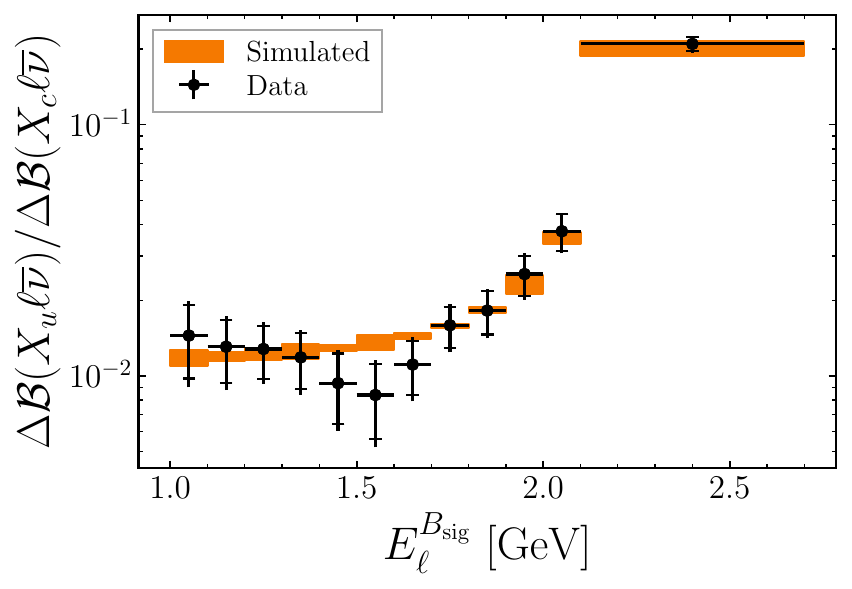} \\
    \includegraphics[width=0.45\textwidth]{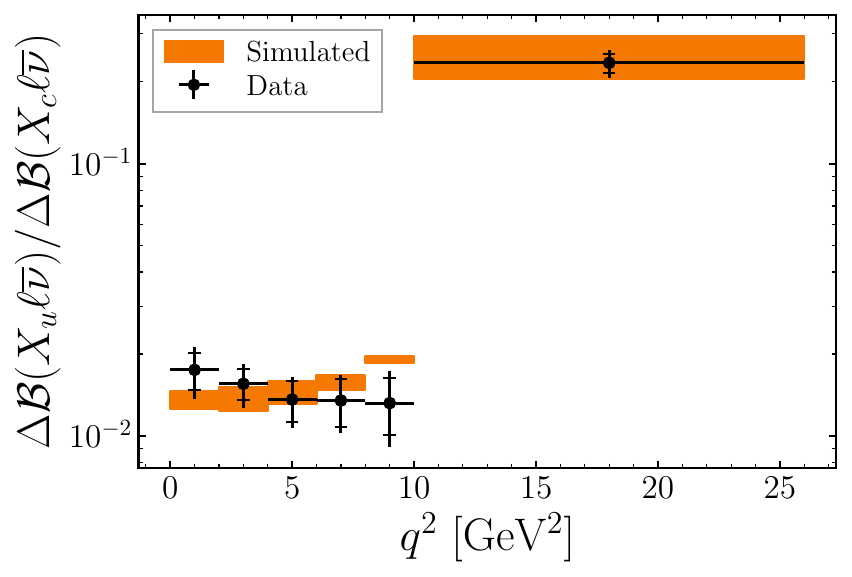} \\
    \caption{Unfolded ratios of partial branching fractions in the phase space $\lepp > 1 \gev$. The simulated unfolded yields have been scaled to the data yields before taking the ratio. The uncertainty on the simulated bands includes all modeling uncertainties discussed in section \ref{sec:syst} and the uncertainty due to the limited size of the MC sample.}
    \label{fig:unfolded_ratios}
\end{figure}

The four signal yields are unfolded using the Singular Value Decomposition (SVD) algorithm of Ref.~\cite{Hocker:1995kb} as implemented in Ref.~\cite{Adye:2011gm,Tackmann:2011np}. The regularization parameter has been tuned to minimize bias from the shape and composition of the $\BXulnu$ and $\BXclnu$ modeling in the nominal phase space. The unfolded yields are then corrected for efficiency and combined to form the ratio of partial branching fractions in the phase space $\lepp>1\gev$. The efficiency-corrected yields of the final three bins in lepton energy, and the final two bins in momentum transfer squared, are summed before taking the ratio. Small biases are observed in the unfolding of the $\BXclnu$ $\lepp$ spectrum near the $1\gev$ threshold. The full size of the bias as estimated from MC simulations is taken as an additional systematic uncertainty. The unfolded differential ratios are presented in Fig. \ref{fig:unfolded_ratios}, the uncertainties are summarized in Tables \ref{tab:unfolded_lepp} and \ref{tab:unfolded_q2}, and the global correlation matrix in Table \ref{tab:unfolding_correlation}. The determination of the ratio of partial branching fractions for a series of increasing thresholds of lepton energy in the $B$ meson rest frame is provided in appendix \ref{sec:appendix_threshold_ratios}.

\begin{table*}
\caption{\label{tab:unfolded_lepp} Summary of the central values ($R$), statistical, and systematic uncertainties for the differential ratio of partial branching fractions as a function of $\lepp$ bin, defined by the boundaries $[1.0,\allowbreak 1.1,\allowbreak 1.2,\allowbreak 1.3,\allowbreak 1.4,\allowbreak 1.5,\allowbreak 1.6,\allowbreak 1.7,\allowbreak 1.8,\allowbreak 1.9,\allowbreak 2.0,\allowbreak 2.1,\allowbreak 2.7] \gev$. The uncertainties are given as relative values on the central value in percent.
} 
\begin{ruledtabular}
\begin{tabular}{l c c c c c c c c c c c c }
$\lepp$ Bin & 0 & 1 &  2 &  3 &  4 &  5 &  6 &  7 &  8 &  9 &  10 &  11 \\ \hline
 $R \times 100$ & $1.45$ & $1.30$ & $1.28$ & $1.18$ & $0.93$ & $0.84$ & $1.11$ & $1.59$ & $1.82$ & $2.54$ & $3.76$ & $21.0$\\  \hline
Stat. Error (Data) & $32.5$ & $28.2$ & $23.9$ & $25.5$ & $31.3$ & $33.0$ & $24.3$ & $18.8$ & $19.7$ & $18.3$ & $16.7$ & $6.3$\\  \hline
$\mathcal{{B}}(\Bbar \to \pi/\eta/\rho/\omega/\eta^\prime \ell \nub)$ & $0.3$ & $0.3$ & $0.3$ & $0.3$ & $0.3$ & $0.3$ & $0.3$ & $0.2$ & $0.2$ & $0.2$ & $0.2$ & $0.2$\\ 
$\mathcal{{FF}}(\Bbar \to \pi/\eta/\rho/\omega/\eta^\prime \ell \nub)$ & $0.3$ & $0.3$ & $0.4$ & $0.4$ & $0.6$ & $0.7$ & $0.6$ & $0.4$ & $0.4$ & $0.3$ & $0.3$ & $0.1$\\ 
$\mathcal{{B}}(\Bbar \rightarrow x_{u} \ell \nub)$ & $1.4$ & $1.4$ & $1.2$ & $1.2$ & $1.3$ & $1.2$ & $0.9$ & $0.7$ & $0.7$ & $0.7$ & $0.5$ & $0.5$\\ 
Hybrid Model (BLNP) & $6.4$ & $2.1$ & $3.3$ & $3.8$ & $1.7$ & $1.2$ & $1.8$ & $1.0$ & $0.5$ & $4.0$ & $0.2$ & $0.2$\\ 
DFN ($m_b^{\rm KN}, a^{\rm KN}$) & $6.6$ & $6.9$ & $7.1$ & $7.6$ & $9.0$ & $10.1$ & $8.2$ & $6.5$ & $5.7$ & $4.3$ & $3.2$ & $0.7$\\ 
$N_{g \to s\overline{s}}$ & $0.3$ & $0.6$ & $0.6$ & $0.8$ & $1.6$ & $2.1$ & $1.3$ & $0.7$ & $0.7$ & $0.4$ & $0.4$ & $0.1$\\  \hline
$\mathcal{{B}}(\Bbar \to D \ell \nub)$ & $0.4$ & $0.4$ & $0.3$ & $0.3$ & $0.2$ & $0.5$ & $0.5$ & $0.4$ & $0.6$ & $0.5$ & $0.2$ & $0.1$\\ 
$\mathcal{{B}}(\Bbar \to D^{*} \ell \nub)$ & $0.3$ & $0.4$ & $0.6$ & $0.7$ & $1.0$ & $1.3$ & $1.2$ & $1.1$ & $1.1$ & $1.2$ & $1.0$ & $0.6$\\ 
$\mathcal{{B}}(\Bbar \to D^{**} \ell \nub)$ & $0.6$ & $0.7$ & $0.8$ & $0.6$ & $0.6$ & $0.5$ & $0.4$ & $1.1$ & $1.3$ & $0.9$ & $0.8$ & $0.2$\\ 
$\mathcal{{B}}(\Bbar \to D^{{(*)}} \eta \ell \nub)$ & $0.9$ & $0.7$ & $0.6$ & $0.9$ & $1.1$ & $0.8$ & $0.5$ & $0.5$ & $0.9$ & $0.8$ & $0.9$ & $0.3$\\ 
$\mathcal{{B}}(\Bbar \to D^{{(*)}} \pi \pi \ell \nub)$ & $0.8$ & $0.8$ & $0.7$ & $0.6$ & $1.1$ & $1.0$ & $0.7$ & $0.5$ & $0.7$ & $0.5$ & $0.6$ & $0.3$\\ 
$\mathcal{FF}(\Bbar \to D \ell \nub)$ & $0.3$ & $0.3$ & $0.3$ & $0.3$ & $0.4$ & $0.4$ & $0.3$ & $0.2$ & $0.2$ & $0.2$ & $0.2$ & $0.2$\\ 
$\mathcal{FF}(\Bbar \to D^{*} \ell \nub)$ & $1.2$ & $1.4$ & $1.5$ & $1.6$ & $2.0$ & $2.3$ & $1.8$ & $1.4$ & $1.3$ & $1.0$ & $0.7$ & $0.8$\\ 
$\mathcal{{FF}}(\Bbar \to D^{**} \ell \nub)$ & $1.3$ & $1.6$ & $1.3$ & $1.1$ & $0.9$ & $0.6$ & $0.7$ & $1.2$ & $1.9$ & $1.9$ & $1.1$ & $0.2$\\ 
Sec.Fakes. Composition & $10.7$ & $9.4$ & $5.3$ & $3.7$ & $5.5$ & $7.8$ & $6.5$ & $4.5$ & $3.9$ & $3.5$ & $3.4$ & $3.9$\\ 
In-situ $q^2$ Calibration & $4.2$ & $4.6$ & $4.5$ & $5.0$ & $6.5$ & $7.6$ & $5.9$ & $4.1$ & $3.8$ & $2.9$ & $2.0$ & $0.6$\\  \hline
$\ell$ID Efficiency & $0.5$ & $0.4$ & $0.4$ & $0.3$ & $0.3$ & $0.2$ & $0.1$ & $0.1$ & $0.2$ & $0.2$ & $0.2$ & $0.1$\\ 
$\ell$ID Fake Rate & $1.9$ & $1.7$ & $1.1$ & $0.7$ & $0.7$ & $0.9$ & $0.8$ & $0.8$ & $0.7$ & $0.6$ & $0.5$ & $0.2$\\ 
$K\pi$ID Efficiency & $3.4$ & $2.8$ & $2.2$ & $2.2$ & $2.8$ & $3.1$ & $2.6$ & $1.9$ & $2.0$ & $1.8$ & $1.6$ & $0.6$\\ 
$K\pi$ID Fake Rate & $3.1$ & $2.2$ & $1.9$ & $2.1$ & $2.4$ & $2.8$ & $2.1$ & $1.5$ & $1.7$ & $1.4$ & $1.4$ & $0.5$\\ 
$K_{S}^{0}$ Efficiency & $0.7$ & $0.6$ & $0.5$ & $0.5$ & $0.7$ & $0.8$ & $0.6$ & $0.5$ & $0.5$ & $0.4$ & $0.4$ & $0.1$\\ 
$\pi_{\rm slow}$ Efficiency & $<0.1$ & $<0.1$ & $<0.1$ & $<0.1$ & $<0.1$ & $<0.1$ & $<0.1$ & $<0.1$ & $<0.1$ & $<0.1$ & $<0.1$ & $<0.1$\\ 
Tracking & $0.2$ & $0.2$ & $0.2$ & $0.1$ & $0.1$ & $<0.1$ & $<0.1$ & $<0.1$ & $0.1$ & $0.1$ & $0.1$ & $0.1$\\ 
Continuum Calibration & $1.5$ & $1.5$ & $1.2$ & $1.0$ & $0.8$ & $0.4$ & $0.3$ & $0.5$ & $0.6$ & $0.4$ & $0.3$ & $0.7$\\ 
$N_{{BB}}$ & $<0.1$ & $<0.1$ & $<0.1$ & $<0.1$ & $<0.1$ & $<0.1$ & $<0.1$ & $<0.1$ & $<0.1$ & $<0.1$ & $<0.1$ & $<0.1$\\ 
$f_{+/0}$ & $0.1$ & $0.1$ & $0.1$ & $0.1$ & $<0.1$ & $<0.1$ & $<0.1$ & $<0.1$ & $<0.1$ & $<0.1$ & $<0.1$ & $<0.1$\\ 
Stat. Error (MC)  & $10.6$ & $9.1$ & $7.5$ & $7.3$ & $8.8$ & $10.5$ & $7.5$ & $5.6$ & $6.0$ & $5.6$ & $4.7$ & $2.0$\\ 
Unfolding Bias & $0.9$ & $2.7$ & $1.8$ & $1.0$ & $0.5$ & $0.4$ & $0.3$ & $0.1$ & $0.2$ & $0.3$ & $<0.1$ & $0.4$\\ \hline
Total Syst. & $19.1$ & $16.8$ & $13.8$ & $13.6$ & $16.2$ & $19.2$ & $15.0$ & $11.3$ & $10.8$ & $10.1$ & $7.6$ & $4.8$\\ 
\end{tabular}
\end{ruledtabular}
\end{table*}

\begin{table*}
\caption{\label{tab:unfolded_q2} Summary of the central values ($R$), statistical, and systematic uncertainties for the differential ratio of partial branching fractions as a function of $q^2$ bin, defined by the boundaries $[0,2,4,6,8,10,26] \gevv$. The uncertainties are given as relative values on the central value in percent.} 
\begin{ruledtabular}
\begin{tabular}{l c c c c c c}
$q^2$ Bin & 0 & 1 &  2 &  3 &  4 &  5 \\ \hline
 $R \times 100$ & $1.75$ & $1.56$ & $1.36$ & $1.35$ & $1.32$ & $23.5$\\  \hline
Stat. Error (Data) & $15.6$ & $13.1$ & $17.0$ & $19.8$ & $23.4$ & $8.0$\\  \hline
$\mathcal{{B}}(\Bbar \to \pi/\eta/\rho/\omega/\eta^\prime \ell \nub)$ & $0.2$ & $0.2$ & $0.2$ & $0.2$ & $0.3$ & $0.2$\\ 
$\mathcal{{FF}}(\Bbar \to \pi/\eta/\rho/\omega/\eta^\prime \ell \nub)$ & $0.2$ & $0.3$ & $0.4$ & $0.4$ & $0.5$ & $0.3$\\ 
$\mathcal{{B}}(\Bbar \rightarrow x_{u} \ell \nub)$ & $1.1$ & $0.9$ & $0.8$ & $0.8$ & $0.8$ & $0.6$\\ 
Hybrid Model (BLNP) & $8.4$ & $7.9$ & $3.8$ & $0.4$ & $2.8$ & $5.5$\\ 
DFN ($m_b^{\rm KN}, a^{\rm KN}$) & $6.5$ & $6.8$ & $6.9$ & $6.5$ & $5.6$ & $2.0$\\ 
$N_{g \to s\overline{s}}$ & $0.4$ & $0.1$ & $0.5$ & $0.7$ & $1.0$ & $0.6$\\  \hline
$\mathcal{{B}}(\Bbar \to D \ell \nub)$ & $0.8$ & $0.5$ & $0.2$ & $0.8$ & $1.2$ & $0.3$\\ 
$\mathcal{{B}}(\Bbar \to D^{*} \ell \nub)$ & $0.2$ & $0.1$ & $0.8$ & $1.6$ & $2.2$ & $1.3$\\ 
$\mathcal{{B}}(\Bbar \to D^{**} \ell \nub)$ & $0.7$ & $0.4$ & $0.5$ & $0.7$ & $0.9$ & $0.3$\\ 
$\mathcal{{B}}(\Bbar \to D^{{(*)}} \eta \ell \nub)$ & $1.9$ & $1.0$ & $0.5$ & $1.1$ & $1.6$ & $0.4$\\ 
$\mathcal{{B}}(\Bbar \to D^{{(*)}} \pi \pi \ell \nub)$ & $0.8$ & $0.7$ & $0.4$ & $0.5$ & $1.2$ & $0.4$\\ 
$\mathcal{FF}(\Bbar \to D \ell \nub)$ & $0.1$ & $0.2$ & $0.4$ & $0.3$ & $0.2$ & $0.1$\\ 
$\mathcal{FF}(\Bbar \to D^{*} \ell \nub)$ & $1.0$ & $1.1$ & $1.6$ & $1.9$ & $4.0$ & $0.7$\\ 
$\mathcal{{FF}}(\Bbar \to D^{**} \ell \nub)$ & $1.9$ & $1.2$ & $1.0$ & $1.3$ & $1.8$ & $0.5$\\ 
Sec.Fakes. Composition & $6.5$ & $4.7$ & $7.9$ & $6.5$ & $6.5$ & $2.9$\\ 
In-situ $q^2$ Calibration & $3.6$ & $3.9$ & $<0.1$ & $6.4$ & $16.0$ & $4.6$\\ \hline
$\ell$ID Efficiency & $0.3$ & $0.1$ & $0.2$ & $0.2$ & $0.2$ & $0.2$\\ 
$\ell$ID Fake Rate & $0.5$ & $0.2$ & $0.6$ & $0.7$ & $0.7$ & $0.2$\\ 
$K\pi$ID Efficiency & $1.9$ & $1.7$ & $1.6$ & $2.4$ & $2.9$ & $0.9$\\ 
$K\pi$ID Fake Rate & $1.5$ & $1.1$ & $1.4$ & $1.6$ & $1.7$ & $0.5$\\ 
$K_{S}^{0}$ Efficiency & $0.6$ & $0.5$ & $0.4$ & $0.7$ & $0.9$ & $0.2$\\ 
$\pi_{\rm slow}$ Efficiency & $<0.1$ & $<0.1$ & $<0.1$ & $<0.1$ & $<0.1$ & $<0.1$\\ 
Tracking & $0.3$ & $0.2$ & $0.1$ & $0.1$ & $0.1$ & $0.1$\\ 
Continuum Calibration & $2.4$ & $0.9$ & $0.7$ & $0.9$ & $0.8$ & $0.3$\\ 
$N_{{BB}}$ & $<0.1$ & $<0.1$ & $<0.1$ & $<0.1$ & $<0.1$ & $<0.1$\\ 
$f_{+/0}$ & $0.1$ & $<0.1$ & $0.1$ & $<0.1$ & $<0.1$ & $<0.1$\\ 
Stat. Error (MC)  & $5.3$ & $4.2$ & $5.5$ & $5.8$ & $6.8$ & $2.3$\\ 
Unfolding Bias & $0.1$ & $<0.1$ & $0.2$ & $0.4$ & $0.4$ & $0.2$\\ \hline
Total Syst. & $14.8$ & $13.2$ & $12.9$ & $13.5$ & $20.7$ & $8.6$\\ 
\end{tabular}
\end{ruledtabular}
\end{table*}

\newcommand{\STAB}[1]{\begin{tabular}{@{}c@{}}#1\end{tabular}}
\begin{table*}
    \centering
    \caption{ Global correlation of the unfolded differential ratio of partial branching fractions in bins of $\lepp$ and $q^2$, defined by the boundaries $[1.0,\allowbreak 1.1,\allowbreak 1.2,\allowbreak 1.3,\allowbreak 1.4,\allowbreak 1.5,\allowbreak 1.6,\allowbreak 1.7,\allowbreak 1.8,\allowbreak 1.9,\allowbreak 2.0,\allowbreak 2.1,\allowbreak 2.7] \gev$ and $[0,2,4,6,8,10,26] \gevv$, respectively. The statistical correlation is given above the diagonal, the systematic below the diagonal.\label{tab:unfolding_correlation}}
    \begin{ruledtabular}
\begin{tabular}{c c | c c c c c c c c c c c c | c c c c c c }
 & &  \multicolumn{12}{c}{\lepp Bin} & \multicolumn{6}{c}{$q^2$ Bin} \\ 
 & & 0 & 1 & 2 & 3 & 4 & 5 & 6 & 7 & 8 & 9 & 10 & 11 & 0 & 1 & 2 & 3 & 4 & 5 \\ \hline 
 \multirow{12}{*}{\STAB{\rotatebox[origin=c]{90}{$\lepp$ Bin}}} &0 & - & $0.87$ & $0.34$ & $-0.09$ & $-0.20$ & $-0.13$ & $-0.05$ & $-0.03$ & $-0.00$ & $0.05$ & $0.09$ & $0.10$ & $0.26$ & $0.20$ & $0.07$ & $0.06$ & $0.07$ & $0.09$ \\ 
 & 1 & $0.92$ & - & $0.68$ & $0.15$ & $-0.13$ & $-0.17$ & $-0.11$ & $-0.04$ & $0.01$ & $0.09$ & $0.13$ & $0.14$ & $0.31$ & $0.27$ & $0.11$ & $0.11$ & $0.14$ & $0.15$ \\ 
 & 2 & $0.71$ & $0.85$ & - & $0.70$ & $0.20$ & $-0.12$ & $-0.17$ & $-0.09$ & $0.03$ & $0.14$ & $0.18$ & $0.19$ & $0.29$ & $0.29$ & $0.14$ & $0.19$ & $0.25$ & $0.21$ \\ 
 & 3 & $0.45$ & $0.54$ & $0.83$ & - & $0.69$ & $0.15$ & $-0.15$ & $-0.14$ & $0.03$ & $0.15$ & $0.19$ & $0.19$ & $0.23$ & $0.25$ & $0.15$ & $0.20$ & $0.28$ & $0.23$ \\ 
 & 4 & $0.17$ & $0.26$ & $0.44$ & $0.75$ & - & $0.64$ & $0.07$ & $-0.15$ & $-0.02$ & $0.14$ & $0.21$ & $0.20$ & $0.18$ & $0.21$ & $0.17$ & $0.21$ & $0.30$ & $0.26$ \\ 
 & 5 & $0.15$ & $0.19$ & $0.34$ & $0.59$ & $0.85$ & - & $0.59$ & $0.04$ & $-0.05$ & $0.11$ & $0.22$ & $0.22$ & $0.18$ & $0.20$ & $0.19$ & $0.24$ & $0.35$ & $0.32$ \\ 
 & 6 & $0.03$ & $0.11$ & $0.25$ & $0.43$ & $0.65$ & $0.85$ & - & $0.59$ & $0.14$ & $0.11$ & $0.20$ & $0.25$ & $0.15$ & $0.19$ & $0.20$ & $0.26$ & $0.40$ & $0.36$ \\ 
 & 7 & $0.04$ & $0.13$ & $0.28$ & $0.42$ & $0.52$ & $0.64$ & $0.86$ & - & $0.63$ & $0.27$ & $0.18$ & $0.24$ & $0.10$ & $0.17$ & $0.20$ & $0.29$ & $0.43$ & $0.40$ \\ 
 & 8 & $0.11$ & $0.16$ & $0.27$ & $0.39$ & $0.46$ & $0.54$ & $0.67$ & $0.88$ & - & $0.69$ & $0.32$ & $0.23$ & $0.12$ & $0.25$ & $0.29$ & $0.37$ & $0.47$ & $0.41$ \\ 
 & 9 & $0.02$ & $0.14$ & $0.20$ & $0.31$ & $0.53$ & $0.51$ & $0.61$ & $0.68$ & $0.80$ & - & $0.68$ & $0.28$ & $0.25$ & $0.46$ & $0.49$ & $0.51$ & $0.52$ & $0.44$ \\ 
 & 10 & $0.16$ & $0.19$ & $0.32$ & $0.45$ & $0.54$ & $0.60$ & $0.59$ & $0.60$ & $0.63$ & $0.74$ & - & $0.48$ & $0.26$ & $0.51$ & $0.51$ & $0.47$ & $0.47$ & $0.42$ \\ 
 & 11 & $0.11$ & $0.14$ & $0.25$ & $0.39$ & $0.48$ & $0.54$ & $0.48$ & $0.42$ & $0.39$ & $0.42$ & $0.63$ & - & $0.29$ & $0.42$ & $0.37$ & $0.34$ & $0.36$ & $0.40$ \\  \hline 
 \multirow{6}{*}{\STAB{\rotatebox[origin=c]{90}{$q^2$ Bin}}} &0 & $0.56$ & $0.45$ & $0.45$ & $0.38$ & $0.06$ & $0.07$ & $-0.08$ & $-0.06$ & $-0.02$ & $-0.20$ & $0.03$ & $-0.07$ & - & $0.43$ & $-0.16$ & $-0.06$ & $0.14$ & $0.23$ \\ 
 & 1 & $0.36$ & $0.26$ & $0.40$ & $0.49$ & $0.28$ & $0.38$ & $0.24$ & $0.24$ & $0.27$ & $0.10$ & $0.43$ & $0.45$ & $0.67$ & - & $0.30$ & $-0.03$ & $0.11$ & $0.25$ \\ 
 & 2 & $0.19$ & $0.19$ & $0.37$ & $0.53$ & $0.49$ & $0.59$ & $0.55$ & $0.57$ & $0.57$ & $0.47$ & $0.69$ & $0.66$ & $0.13$ & $0.70$ & - & $0.49$ & $0.14$ & $0.20$ \\ 
 & 3 & $0.20$ & $0.26$ & $0.41$ & $0.55$ & $0.63$ & $0.70$ & $0.74$ & $0.76$ & $0.72$ & $0.67$ & $0.73$ & $0.59$ & $-0.15$ & $0.23$ & $0.71$ & - & $0.57$ & $0.22$ \\ 
 & 4 & $0.16$ & $0.27$ & $0.38$ & $0.49$ & $0.62$ & $0.66$ & $0.71$ & $0.70$ & $0.66$ & $0.65$ & $0.58$ & $0.43$ & $-0.29$ & $-0.11$ & $0.31$ & $0.80$ & - & $0.38$ \\ 
 & 5 & $-0.00$ & $0.19$ & $0.23$ & $0.30$ & $0.60$ & $0.53$ & $0.63$ & $0.58$ & $0.52$ & $0.72$ & $0.47$ & $0.45$ & $-0.45$ & $-0.27$ & $0.18$ & $0.56$ & $0.74$ & - \\ 
\end{tabular}
\end{ruledtabular}
\end{table*}

\section{\label{sec:summary} Summary and Conclusion}

We report the first measurements of the ratio of partial branching fractions of the inclusive charmless to the inclusive charmed semileptonic $B$ decay using the full Belle data-set of $711\invfb$ and state-of-the-art software developed for \belletwo. The $\BXulnu$ yield is extracted using a data-driven description of $\BXclnu$ decays to minimize potential bias due to mismodeling of this component, which dominates $\BXulnu$. In the region $E_\ell^{B_{\rm sig}} > 1.0 \gev$, we measure:
\begin{equation}
    \frac{\Delta \mathcal{B}(\BXulnu)}{\Delta \mathcal{B}(\BXclnu)} = (1.99 \pm 0.17_{\rm stat} \pm 0.16_{\rm syst}) \times 10^{-2},
\end{equation}
where the uncertainties are statistical and systematic, respectively. From the partial branching fraction ratio we extract $\frac{|\Vub|}{|\Vcb|}$ using two theoretical calculations for the partial decay rate of $\BXulnu$, finding 

\begin{align}
    \frac{|\Vub|}{|\Vcb|}^{\rm BLNP} &= (9.81 \pm 0.42_{\rm stat} \pm 0.38_{\rm syst}  \nonumber \\
        &{\hphantom{}} \pm 0.51_{ \Delta\Gamma(\BXulnu)} \pm 0.20_{ \Delta\Gamma(\BXclnu)}) \times 10^{-2}, \\
    \frac{|\Vub|}{|\Vcb|}^{\rm GGOU} &= (10.06 \pm 0.43_{\rm stat} \pm 0.39_{\rm syst} \nonumber \\
        &{\hphantom{}} \pm 0.23_{ \Delta\Gamma(\BXulnu)} \pm 0.20_{ \Delta\Gamma(\BXclnu)}) \times 10^{-2},
\end{align}

where the third and fourth uncertainties are from the partial decay rates of $\BXulnu$ and $\BXclnu$, respectively. These values are in excellent agreement with inclusive and exclusive world averages of $|\Vub|/|\Vcb|$.
We additionally report this measurement broken down by $B$ charge, determining a limit on isospin breaking in $\BXulnu$ decays, as well as by lepton flavor, finding no evidence for lepton flavor violation in either the charmed or charmless inclusive semileptonic $B$ decay. Furthermore, we report the differential ratio of partial branching fractions in $q^2$ and $\lepp$. 
\section{\label{sec:ack} Acknowledgments}
This work, based on data collected using the Belle detector, which was
operated until June 2010, was supported by 
the Ministry of Education, Culture, Sports, Science, and
Technology (MEXT) of Japan, the Japan Society for the 
Promotion of Science (JSPS), and the Tau-Lepton Physics 
Research Center of Nagoya University; 
the Australian Research Council including grants
DP180102629, 
DP170102389, 
DP170102204, 
DE220100462, 
DP150103061, 
FT130100303; 
Austrian Federal Ministry of Education, Science and Research (FWF) and
FWF Austrian Science Fund No.~P~31361-N36;
the National Natural Science Foundation of China under Contracts
No.~11675166,  
No.~11705209;  
No.~11975076;  
No.~12135005;  
No.~12175041;  
No.~12161141008; 
Key Research Program of Frontier Sciences, Chinese Academy of Sciences (CAS), Grant No.~QYZDJ-SSW-SLH011; 
Project ZR2022JQ02 supported by Shandong Provincial Natural Science Foundation;
the Ministry of Education, Youth and Sports of the Czech
Republic under Contract No.~LTT17020;
the Czech Science Foundation Grant No. 22-18469S;
Horizon 2020 ERC Advanced Grant No.~884719 and ERC Starting Grant No.~947006 ``InterLeptons'' (European Union);
the Carl Zeiss Foundation, the Deutsche Forschungsgemeinschaft, the
Excellence Cluster Universe, and the VolkswagenStiftung;
the Department of Atomic Energy (Project Identification No. RTI 4002) and the Department of Science and Technology of India; 
the Istituto Nazionale di Fisica Nucleare of Italy; 
National Research Foundation (NRF) of Korea Grant
Nos.~2016R1\-D1A1B\-02012900, 2018R1\-A2B\-3003643,
2018R1\-A6A1A\-06024970, RS\-2022\-00197659,
2019R1\-I1A3A\-01058933, 2021R1\-A6A1A\-03043957,
2021R1\-F1A\-1060423, 2021R1\-F1A\-1064008, 2022R1\-A2C\-1003993;
Radiation Science Research Institute, Foreign Large-size Research Facility Application Supporting project, the Global Science Experimental Data Hub Center of the Korea Institute of Science and Technology Information and KREONET/GLORIAD;
the Polish Ministry of Science and Higher Education and 
the National Science Center;
the Ministry of Science and Higher Education of the Russian Federation, Agreement 14.W03.31.0026, 
and the HSE University Basic Research Program, Moscow; 
University of Tabuk research grants
S-1440-0321, S-0256-1438, and S-0280-1439 (Saudi Arabia);
the Slovenian Research Agency Grant Nos. J1-9124 and P1-0135;
Ikerbasque, Basque Foundation for Science, Spain;
the Swiss National Science Foundation; 
the Ministry of Education and the Ministry of Science and Technology of Taiwan;
and the United States Department of Energy and the National Science Foundation.
These acknowledgments are not to be interpreted as an endorsement of any
statement made by any of our institutes, funding agencies, governments, or
their representatives.
We thank the KEKB group for the excellent operation of the
accelerator; the KEK cryogenics group for the efficient
operation of the solenoid; and the KEK computer group and the Pacific Northwest National
Laboratory (PNNL) Environmental Molecular Sciences Laboratory (EMSL)
computing group for strong computing support; and the National
Institute of Informatics, and Science Information NETwork 6 (SINET6) for
valuable network support.

\appendix
\section{}
\label{sec:appendix_threshold_ratios}

From the unfolded yields we calculate the ratio of partial branching fractions as a function of the lepton energy threshold as: 

\begin{equation}
    \frac{\Delta \mathcal{B}(\BXulnu)}{\Delta \mathcal{B}(\BXclnu)}|_{\lepp > Y} = \frac{\sum_{i > y} N_{i,{ \rm unf.}}^{\Xulnu} / \epsilon^{\Xulnu}_i}{\sum_{i > y} N_{i,{ \rm unf.}}^{\Xclnu} / \epsilon^{\Xclnu}_i}
\end{equation}
where $Y$ is a threshold corresponding to the lower bin boundary of the $y$-th bin, $N_{i,{ \rm unf.}}^{\BXuclnu}$, are the unfolded yields and $\epsilon^{\BXuclnu}_i$ are the reconstruction efficiencies in the $i$-th bin of the \BXulnu and \BXclnu samples, respectively. The unfolded ratios of partial branching fractions as a function of lepton emergy are summarized in Table \ref{tab:unfolded_lepp_threshhold} and the global correlation matrix in Table \ref{tab:unfolded_lepp_threshhold_correlation}. As a cross-check we perform a direct fit for each lepton energy threshold following the procedure established in section \ref{sec:measurement}. The fit ratios are consistent with the unfolded ratios within $2.1\sigma$.

\begin{table*}
\caption{\label{tab:unfolded_lepp_threshhold} Summary of the central values ($R$), statistical, and systematic uncertainties for the ratio of partial branching fractions as a function of increasing $\lepp$ threshold. The uncertainties are given as relative values on the central value in percent.}
\begin{ruledtabular}
\begin{tabular}{l c c c c c c c c c c c c }
$\lepp$ Threshold $[\gev]$ & $1.0$  & $1.1$  & $1.2$  & $1.3$  & $1.4$  & $1.5$  & $1.6$  & $1.7$  & $1.8$  & $1.9$  & $2.0$  \\ \hline 
$R \times100$ & $1.94$ & $1.98$ & $2.04$ & $2.13$ & $2.27$ & $2.51$ & $2.90$ & $3.47$ & $4.29$ & $5.87$ & $9.29$\\ \hline
Stat. Error (Data) & $8.9$ & $9.0$ & $9.3$ & $9.6$ & $9.8$ & $9.7$ & $9.5$ & $9.7$ & $9.6$ & $8.7$ & $7.7$\\ \hline
$\mathcal{{B}}(\Bbar \to \pi/\eta/\rho/\omega/\eta^\prime \ell \nub)$ & $0.2$ & $0.2$ & $0.2$ & $0.2$ & $0.2$ & $0.2$ & $0.2$ & $0.2$ & $0.2$ & $0.2$ & $0.2$\\ 
$\mathcal{{FF}}(\Bbar \to \pi/\eta/\rho/\omega/\eta^\prime \ell \nub)$ & $0.3$ & $0.3$ & $0.3$ & $0.3$ & $0.3$ & $0.3$ & $0.3$ & $0.3$ & $0.2$ & $0.2$ & $0.1$\\ 
$\mathcal{{B}}(B \rightarrow x_{u} \ell \nu)$ & $0.8$ & $0.8$ & $0.8$ & $0.7$ & $0.7$ & $0.7$ & $0.6$ & $0.6$ & $0.6$ & $0.6$ & $0.5$\\ 
Hybrid Model (BLNP) & $0.4$ & $<0.1$ & $0.1$ & $0.3$ & $0.7$ & $0.6$ & $0.7$ & $0.6$ & $0.6$ & $0.7$ & $0.2$\\ 
DFN ($m_b^{\rm KN}, a^{\rm KN}$) & $4.9$ & $4.8$ & $4.7$ & $4.6$ & $4.3$ & $4.0$ & $3.6$ & $3.2$ & $2.7$ & $2.1$ & $1.4$\\ 
$N_{s\overline{s}}$ & $0.6$ & $0.6$ & $0.6$ & $0.6$ & $0.6$ & $0.5$ & $0.4$ & $0.3$ & $0.3$ & $0.2$ & $0.2$\\ \hline
$\mathcal{{B}}(B \rightarrow D \ell \nu)$ & $0.1$ & $0.1$ & $0.2$ & $0.2$ & $0.2$ & $0.2$ & $0.2$ & $0.2$ & $0.2$ & $0.1$ & $0.1$\\ 
$\mathcal{{B}}(B \rightarrow D^{*} \ell \nu)$ & $0.8$ & $0.8$ & $0.9$ & $0.9$ & $0.9$ & $0.9$ & $0.9$ & $0.9$ & $0.9$ & $0.8$ & $0.7$\\ 
$\mathcal{{B}}(B \rightarrow D^{**} \ell \nu)$ & $0.3$ & $0.3$ & $0.3$ & $0.4$ & $0.5$ & $0.5$ & $0.6$ & $0.6$ & $0.6$ & $0.5$ & $0.4$\\ 
$\mathcal{{B}}(B \rightarrow D^{{(*)}} \eta \ell \nu)$ & $0.2$ & $0.1$ & $0.1$ & $0.1$ & $0.2$ & $0.3$ & $0.4$ & $0.4$ & $0.5$ & $0.5$ & $0.4$\\ 
$\mathcal{{B}}(B \rightarrow D^{{(*)}} \pi \pi \ell \nu)$ & $0.2$ & $0.2$ & $0.2$ & $0.2$ & $0.2$ & $0.3$ & $0.3$ & $0.3$ & $0.3$ & $0.3$ & $0.3$\\ 
$\mathcal{FF}(B \rightarrow D \ell \nu)$ & $0.2$ & $0.2$ & $0.2$ & $0.2$ & $0.2$ & $0.2$ & $0.2$ & $0.2$ & $0.1$ & $0.1$ & $0.1$\\ 
$\mathcal{FF}(B \rightarrow D^{*} \ell \nu)$ & $1.1$ & $1.1$ & $1.0$ & $1.0$ & $1.0$ & $0.9$ & $0.8$ & $0.7$ & $0.7$ & $0.6$ & $0.5$\\ 
$\mathcal{{FF}}(B \rightarrow D^{**} \ell \nu)$ & $0.3$ & $0.4$ & $0.4$ & $0.6$ & $0.7$ & $0.8$ & $0.9$ & $0.9$ & $0.9$ & $0.7$ & $0.4$\\ 
Sec.Fakes. Composition & $3.6$ & $3.7$ & $3.9$ & $4.1$ & $4.2$ & $4.1$ & $3.8$ & $3.6$ & $3.6$ & $3.6$ & $3.7$\\ 
In-situ $q^2$ Calibration & $3.3$ & $3.3$ & $3.2$ & $3.1$ & $3.0$ & $2.7$ & $2.4$ & $2.1$ & $1.7$ & $1.3$ & $0.9$\\  \hline
$\ell$ID Efficiency & $0.1$ & $0.1$ & $0.1$ & $0.1$ & $0.1$ & $0.1$ & $0.1$ & $0.1$ & $0.1$ & $0.1$ & $0.1$\\ 
$\ell$ID Fake Rate & $0.2$ & $0.2$ & $0.3$ & $0.4$ & $0.5$ & $0.5$ & $0.5$ & $0.5$ & $0.4$ & $0.4$ & $0.3$\\ 
$K\pi$ID Efficiency & $1.1$ & $1.1$ & $1.1$ & $1.1$ & $1.1$ & $1.1$ & $1.1$ & $1.1$ & $1.0$ & $0.9$ & $0.8$\\ 
$K\pi$ID Fake Rate & $0.5$ & $0.5$ & $0.6$ & $0.6$ & $0.6$ & $0.6$ & $0.7$ & $0.7$ & $0.7$ & $0.7$ & $0.6$\\ 
$K_{S}^{0}$ Efficiency & $0.2$ & $0.2$ & $0.2$ & $0.2$ & $0.2$ & $0.2$ & $0.2$ & $0.2$ & $0.2$ & $0.2$ & $0.2$\\ 
$\pi_{\rm slow}$ Efficiency & $<0.1$ & $<0.1$ & $<0.1$ & $<0.1$ & $<0.1$ & $<0.1$ & $<0.1$ & $<0.1$ & $<0.1$ & $<0.1$ & $<0.1$ \\ 
Tracking & $0.1$ & $0.1$ & $0.1$ & $0.1$ & $0.1$ & $0.1$ & $0.1$ & $0.1$ & $0.1$ & $0.1$ & $0.1$\\ 
Continuum Calibration & $0.4$ & $0.4$ & $0.3$ & $0.3$ & $0.3$ & $0.2$ & $0.2$ & $0.3$ & $0.3$ & $0.4$ & $0.6$\\ 
$N_{{BB}}$ & $<0.1$ & $<0.1$ & $<0.1$ & $<0.1$ & $<0.1$ & $<0.1$ & $<0.1$ & $<0.1$ & $<0.1$ & $<0.1$ & $<0.1$ \\ 
$f_{+/0}$ & $<0.1$ & $<0.1$ & $<0.1$ & $<0.1$ & $<0.1$ & $<0.1$ & $<0.1$ & $<0.1$ & $<0.1$ & $<0.1$ & $<0.1$ \\ 
Stat. Error (MC)   & $2.8$ & $2.8$ & $2.8$ & $2.8$ & $2.9$ & $2.8$ & $2.8$ & $2.9$ & $2.9$ & $2.6$ & $2.2$\\ 
Unfolding Bias & $0.6$ & $0.7$ & $0.5$ & $0.4$ & $0.3$ & $0.2$ & $0.2$ & $0.1$ & $0.1$ & $<0.1$ & $0.2$\\ \hline
Total Syst. & $7.9$ & $7.8$ & $7.8$ & $7.8$ & $7.7$ & $7.4$ & $6.9$ & $6.4$ & $6.0$ & $5.5$ & $5.0$\\ 
\end{tabular}
\end{ruledtabular}
\end{table*}

\begin{table*}
    \centering
    \caption{\label{tab:unfolded_lepp_threshhold_correlation} Global correlation of the ratio of partial branching fractions as a function of $\lepp$ thresholds. The statistical correlation is given above the diagonal, the systematic below the diagonal.}
    \begin{ruledtabular}
\begin{tabular}{c c | c c c c c c c c c c c }
 & &  \multicolumn{11}{c}{$\lepp$ Threshold $[\gev]$} \\
 & & $1.0$  & $1.1$  & $1.2$  & $1.3$  & $1.4$  & $1.5$  & $1.6$  & $1.7$  & $1.8$  & $1.9$  & $2.0$  \\ \hline 
 \multirow{12}{*}{\STAB{\rotatebox[origin=c]{90}{$\lepp$ Bin}}} &$1.0$  & - & $0.98$ & $0.94$ & $0.90$ & $0.86$ & $0.83$ & $0.81$ & $0.79$ & $0.79$ & $0.78$ & $0.72$ \\ 
 & $1.1$  & $0.99$ & - & $0.99$ & $0.95$ & $0.91$ & $0.87$ & $0.84$ & $0.82$ & $0.82$ & $0.80$ & $0.74$ \\ 
 & $1.2$  & $0.97$ & $0.99$ & - & $0.99$ & $0.95$ & $0.90$ & $0.86$ & $0.84$ & $0.83$ & $0.81$ & $0.74$ \\ 
 & $1.3$  & $0.95$ & $0.98$ & $0.99$ & - & $0.98$ & $0.94$ & $0.89$ & $0.86$ & $0.84$ & $0.81$ & $0.74$ \\ 
 & $1.4$  & $0.92$ & $0.96$ & $0.98$ & $0.99$ & - & $0.98$ & $0.94$ & $0.89$ & $0.86$ & $0.82$ & $0.74$ \\ 
 & $1.5$  & $0.91$ & $0.94$ & $0.97$ & $0.98$ & $1.00$ & - & $0.98$ & $0.93$ & $0.88$ & $0.83$ & $0.74$ \\ 
 & $1.6$  & $0.89$ & $0.92$ & $0.95$ & $0.97$ & $0.98$ & $0.99$ & - & $0.98$ & $0.92$ & $0.85$ & $0.75$ \\ 
 & $1.7$  & $0.87$ & $0.91$ & $0.93$ & $0.95$ & $0.96$ & $0.98$ & $0.99$ & - & $0.97$ & $0.89$ & $0.77$ \\ 
 & $1.8$  & $0.85$ & $0.89$ & $0.91$ & $0.92$ & $0.94$ & $0.95$ & $0.97$ & $0.99$ & - & $0.96$ & $0.83$ \\ 
 & $1.9$  & $0.82$ & $0.85$ & $0.87$ & $0.89$ & $0.90$ & $0.91$ & $0.92$ & $0.94$ & $0.98$ & - & $0.93$ \\ 
 & $2.0$  & $0.76$ & $0.78$ & $0.80$ & $0.81$ & $0.82$ & $0.82$ & $0.83$ & $0.85$ & $0.89$ & $0.95$ & - \\ 
\end{tabular}

\end{ruledtabular}
\end{table*}

\bibliography{references}

\end{document}